\documentclass[a4paper]{jpconf}
\pdfoutput=1

\usepackage[utf8x]{inputenc}
\usepackage{amsmath}
\usepackage{amssymb}
\usepackage{bbm}
\usepackage{graphicx}
\usepackage{float}

\begin{document}

\noindent
http://dx.doi.org/10.1088/1742-6596/343/1/012105

\title{$\kappa$-Poincar\'e phase space: speed of massless particles and relativity of spacetime locality}

\author{G.~ROSATI$^{1,2}$, N.~LORET$^{1,2}$, G.~AMELINO-CAMELIA$^{1,2}$}
\address{$^1$ Dipartimento di Fisica, Universit\`a di Roma ``La Sapienza", P.le A. Moro 2, 00185 Roma, EU}
\address{$^2$ INFN, Sez.~Roma1, P.le A. Moro 2, 00185 Roma, EU}

\begin{abstract}
The study of phase-space constructions based on the properties of the $\kappa$-Poincar\'e Hopf
algebra has been a very active area, mostly because of its possible applications in the
phenomenology of Planck-scale-induced momentum dependence of the speed of ultrarelativistic particles.
We here summarize, with emphasis on the emerging relativity of spacetime locality, 
some results relevant for this research program that were recently reported
in arXiv:1006.2126 (PhysRevLett106,071301)
 arXiv:1102.4637 (PhysLettB700,150)
and arXiv:1107.1724
\end{abstract}

\section{Introduction}
Over the last decade there has been a strong effort~\cite{gacLRR}
 aimed at seeking experimental evidence of Planck-scale ($M_p \sim 10^{19} GeV$) effects which could be motivated from the study of the quantum-gravity problem.
 One of the most studied opportunities concerns the possibility that the speed of massless particles (photons) might have a Planck-scale-induced dependence on wavelength/momentum,
 as suggested by several preliminary studies in some of the alternative directions
 of research on quantum gravity (see, {\it e.g.}, Refs.~\cite{grbgac,gampul,mexweave,gacmaj}). 
 This is of particular interest in the context of observations of gamma-ray-bursts
 where some of the possible scenarios for the form of this momentum (and Planck-scale) dependence
  could have observably-large manifestations~\cite{gacLRR,grbgac,gampul,schaefer,fermiSCIENCE,ellisUNO,gacsmolinPRD,fermiNATURE}.
 
From the conceptual perspective the results and models that shaped this research line
have been primarily analyzed for what concerns the fate of Lorentz symmetry.
A momentum-dependent speed of photons (and other particles, when their mass can be neglected)
can evidently be accommodated in scenarios in which Lorentz symmetry is ``broken", violating
the relativity of inertial frames~\cite{grbgac,gampul} (all see, for cases however without
momentum dependence of the effect, Refs.~\cite{kosteSMEfirst,colglafirst}).
Starting with the proposal put forward in Refs.~\cite{dsr1,dsr2} there has also been 
growing interest (also see, {\it e.g.}, 
Refs.~\cite{jurekdsr1,dsrnature,leeDSRprd,leeDSRrainbow,jurekdsrREVIEW,gacdsrrev2010}
and references therein)
in the possibility that Lorentz symmetry might be ``deformed", preserving the relativity
of inertial frames at the ``cost" of modifying the laws of transformation between observers.

We here focus on this deformed-Lorentz-symmetry scenario and particularly on the
possibility that some viable phenomenological models of this type could be 
inspired~\cite{dsr1,jurekdsrREVIEW,gacdsrrev2010}
by the mathematical structure of 
the $\kappa$-Poincar\'e Hopf algebra\cite{lukie1992,majrue,lukieANNALS}.
Specifically, we here summarize the results of the studies
originally reported in Refs.~\cite{bob,k-bob,anatomy}
where phase-space constructions inspired by $\kappa$-Poincar\'e 
were used to formalize the possibility
of dependence on momentum of the speed of massless particles 
which in leading order are linear:
\begin{equation}
v_{m=0} \simeq 1- \ell |{\bf p}| ~,
\label{velocityDSR}
\end{equation}
where $\ell$ is an inverse-momentum deformation scale 
(usually thought to be roughly of the order of $|\ell| \sim 1/M_p$) 
which can take both positive or negative values (in the sense that one can contemplate 
both models with $\ell M_p > 0$ and models
 with $\ell M_p < 0$), and it is assumed that (\ref{velocityDSR}) would hold 
 for $|{\bf p}| \ll \ell^{-1}$.
 
 Besides the momentum dependence of the speed of photons we shall also put in focus
 the implications that the relevant deformations of Lorentz symmetry have
 for the relativity of spacetime locality~\cite{bob,k-bob,leeINERTIALlimit}.
 We shall do this first, on the basis of Refs.~\cite{bob,k-bob}, in setups 
 which are confined to the description of free particles,
 and then, in the part that summarizes the findings of Ref.~\cite{anatomy},
 we shall describe interacting particles within the ``relative-locality framework"
 proposed in Ref.~\cite{prl}, which is centered on the geometry of momentum space.

We work throughout at leading order in the deformation scale.
This keeps formulas at reasonably manageable level, sufficiently characterizes the
new concepts, and would be fully sufficient for phenomenology
if indeed the deformation scale  is roughly of the order
of the huge Planck scale (in which case a leading-order analysis should be all we
need for comparison to data we could realistically imagine to gather
over the next few decades).\\
Some of the results are also discussed specifically
for 1+1-dimensional cases, where all conceptual issues
here relevant are already present and can be exposed more simply.

\section{Deformations of Lorentz symmetry}\label{dsrgeneral}
The notion of deformed Lorentz symmetry which we shall here adopt is the one
of the proposal ``DSR" (``doubly-special", or ``deformed-special", relativity),
first introduced in Refs.~\cite{dsr1,dsr2}).\\
This proposal was put forward
as a possible
description of {\underline{preliminary}}
theory results suggesting that there {\underline{might}} be violations
of some special-relativistic laws in certain approaches to the quantum-gravity problem,
most notably the ones based on spacetime noncommutativity and loop quantum gravity.
The part of the quantum-gravity community interested in those results was interpreting them
as a manifestation of a full breakdown of Lorentz symmetry, with the emergence of
a preferred class of observers (an ``ether"). But it was argued in Ref.~\cite{dsr1}
that departures from Special Relativity governed by a high-energy/short-distance scale
may well be compatible with the Relativity Principle, the principle of relativity
of inertial observers, at the cost of allowing some consistent modifications
of relativistic kinematics and 
of the Poincar\'e/Lorentz transformations.

The main area of investigation of the DSR proposal has been for the last decade
the possibility of introducing relativistically some Planck-scale-deformed on-shell relations.
The  DSR proposal was put forward~\cite{dsr1} as a conceptual path for pursuing
a broader class of scenarios of interest for fundamental physics, 
including the possibility of introducing the second
observer-independent scale primitively in spacetime structure or primitively at the
level of the (deformed) de Broglie relation between wavelength and momentum.
However, the bulk of the preliminary results providing encouragement for this
approach came from quantum-gravity research
concerning Planck-scale departures from the special-relativistic on-shell relation,
and this in turn became the main focus of DSR research.

This idea of deformed Lorentz symmetry is actually very simple, as we shall here render manifest
on the basis of an analogy with how the Poincar\'e transformations came to be adopted
as a deformation of Galileo transformations.
Famously, as the Maxwell formulation of electromagnetism,
with an observer-independent speed scale ``$c$", gained more and more
experimental support (among which we count the Michelson-Morley results)
it became clear that Galilean relativistic symmetries could no longer be upheld.
From a modern perspective we should see the pre-Einsteinian attempts to address that
crisis (such as the ones of Lorentz) as attempts to ``break Galilean invariance",
 {\it i.e.} preserve the validity of Galilean transformations
as laws of transformation among inertial observers, but renouncing to the possibility that those
transformations be a symmetry of the laws of physics. The ``ether" would be a preferred frame
for the description of the laws of physics, and the laws of physics that hold in other frames
would be obtained from the ones of the preferred frame via Galilean transformations.\\
Those attempts failed.
What succeeded is completely complementary. Experimental evidence, and the analyses
of Einstein (and Poincar\'e) led us to a ``deformation of Galilean invariance":
in Special Relativity the laws of transformation
among observers still are a symmetry of the laws of physics (Special Relativity is no less
relativistic than Galilean Relativity), but the special-relativistic transformation laws
are a $c$-deformation of the Galilean laws of transformation with the special property
of achieving the observer-independence of the speed scale $c$.

This famous $c$-deformation in particular replaces the Galilean on-shell relation
$E= \mathrm{constant} + {\bf p}^2/(2m)$ with the special-relativistic version
$$E= \sqrt{c^2{\bf p}^2+c^4 m^2}~,$$
and the Galilean
composition of velocities ${\bf u} \oplus {\bf v} = {\bf u} + {\bf v}$
with the special relativistic law of composition of velocities
\begin{equation}
\!\!\!\!\!\!\!\!\!\!\!\!\!\!
{\bf u} \oplus_c {\bf v} = \frac{1}{1+\frac{{\bf u} \cdot {\bf v}}{c^2}}
\left({\bf u} + \frac{1}{\gamma_u}  {\bf v}
+ \frac{1}{c^2} \frac{\gamma_u}{1+\gamma_u} ({\bf u} \cdot {\bf v}){\bf u}  \right)
\label{ungarVEL}
\end{equation}
where as usual $\gamma_u \equiv 1/\sqrt{1- {\bf u} \cdot {\bf u}/c^2}$.\\
The richness of the velocity-composition (\ref{ungarVEL})
is a necessary match for the demanding task of introducing an absolute scale in a relativistic theory.
And it is unfortunate that undergraduate textbooks often choose to limit the discussion to the
special case of (\ref{ungarVEL}) which applies when ${\bf u}$ and ${\bf v}$ are collinear:
\begin{equation}
{\bf u} \oplus_c {\bf v} \Big|_{collinear}
= \frac{{\bf u} + {\bf v}}{1+\frac{{\bf u} \cdot {\bf v}}{c^2}}~.
\label{velTextbook}
\end{equation}
The invariance of the velocity scale $c$ of course requires that boosts act non-linearly on
velocity space, and this is visible not only in (\ref{ungarVEL}) but also in (\ref{velTextbook}).
But also the non-commutativity and non-associativity
 of (\ref{ungarVEL}) (which are silenced in (\ref{velTextbook}))
play a central role~\cite{ungar,ungarFOLLOWER,florianeteraUNGAR}
 in the logical consistency of Special Relativity as a theory
enforcing relativistically the absoluteness of the speed scale $c$.
For example, the composition law (\ref{ungarVEL}) encodes Thomas-Wigner rotations,
and in turn the relativity of simultaneity.

Equipped with this quick reminder of some features of the transition from
Galilean Relativity to Special Relativity we can now quickly summarize
the logical ingredients of a DSR framework. The analogy is particularly close in cases
where the DSR-deformation of Lorentz symmetry is introduced primitively
at the level of the on-shell relation.
To see this let us consider an on-shell relation 
\begin{equation}
m^2 = p_0^2 - {\bf p}^2 + \Delta(E,{\bf p};\ell)
\label{dsr1gen}
\end{equation}
where $\Delta$ is the deformation and $\ell$ is the deformation scale.

Evidently when $\Delta \neq 0$ such an on-shell relation (\ref{dsr1gen})
is not Lorentz invariant. If we insist on this law and on
the validity of classical (undeformed) Lorentz transformations between inertial
observers we clearly end up with a preferred-frame picture, and the Principle
of Relativity of inertial frames must be abandoned: the scale $\ell$ cannot
be observer independent, and actually the whole form of (\ref{dsr1gen}) is subject
to vary from one class of inertial observers to another.\\
The other option~\cite{dsr1} in such cases is the DSR option of enforcing
the relativistic invariance of (\ref{dsr1gen}), preserving the relativity
of inertial frames, at the cost of modifying the action of boosts on momenta.
Then in such theories both the velocity scale $c$ (here mute only because of the
choice of dimensions) and the energy scale $\ell$ play the
same role~\cite{dsr1,dsrnature}
of invariant scales of the relativistic theory which govern the form of boost
transformations. \\
Several examples of boost deformations adapted in the DSR sense to modified on-shell
relations have been analyzed in some detail
(see {\it e.g.} Refs.~\cite{jurekdsr1,dsrnature,leeDSRprd,leeDSRrainbow,jurekdsrREVIEW,gacdsrrev2010,goldenrule}).
Clearly these DSR-deformed boosts ${\cal N}_j$ must be such that
they admit the deformed shell, $p_0^2 - {\bf p}^2 + \Delta(E,{\bf p};M_*)$,
as an invariant,
and in turn the law of composition of momenta must also be
deformed~\cite{dsr1}, $p_\mu \oplus_{\cal N} k_\mu$,
since it must be covariant~\cite{dsr1,goldenrule} 
under the action of the (DSR-deformed) boost transformations.\\
All this is evidently analogous to corresponding aspects of Galilean Relativity and Special Relativity:
of course in all these cases the on-shell relation 
is boost invariant (but respectively under Galilean boosts,
Lorentz boosts, and DSR-deformed Lorentz boosts); for Special Relativity the action of boosts
evidently must depend on the speed scale $c$ and must act non-linearly on velocities
(since it must enforce observer independence of $c$-dependent laws), and for DSR relativity
the action of boosts
evidently must depend on both the speed scale $c$ and the scale $\ell$, acting non-linearly
both on velocities and momenta, since it must enforce observer independence of $c$-dependent and $\ell$-dependent laws (the scale $\ell$ is endowed in DSR with properties that are completely
analogous to the familiar properties of $c$; DSR-relativistic theories have 
both $c$ and $\ell$ as relativistic-invariant
scales).

\section{$\kappa$-Poincar\'e phase-space construction with Minkowski coordinates}
Let us start, in this section, by summarizing the main results of Ref.~\cite{bob},
where a relativistic description of momentum dependence of the speed of massless particles
was given within a classical-phase-space construction inspired by
properties of the $\kappa$-Poincar\'e Hopf algebra~\cite{lukie1992,majrue,lukieANNALS}.\\
At the quantum level the $\kappa$-Poincar\'e Hopf algebra is intimately
related to the $\kappa$-Minkowski noncommutative spacetime~\cite{majrue,lukieANNALS},
on which we shall return later. But for classical-phase-space constructions one can
consider standard Minkowski spacetime coordinates in combination with
a description of relativistic transformations  inspired by
properties of the $\kappa$-Poincar\'e Hopf algebra.
So in this section we assume trivial Poisson brackets for
the spacetime coordinates, $\{ x , t \} = 0$.

Actually Ref.~\cite{bob} analyzed a 3-parameter family of phase spaces
suitable for a DSR-relativistic implementation of the speed law (\ref{velocityDSR}). 
But in order to highlight the possible connection with the $\kappa$-Poincar\'e Hopf algebra
we here specialize to the case involving the following
description of the boost and translation generators in a 1+1-dimensional context:
\begin{equation}
\left\{ E ,p\right\} =0 \, ,~\,
 \left\{ \mathcal{N}, E \right\} = p \, ,~\,
 \left\{ \mathcal{N},p \right\} =E + \ell E^{2} + \frac{\ell}{2}p^{2} \ ,
\label{poinca} \\
\end{equation}
which is inspired by
the so-called ``$\kappa$-Poincar\'e bicrossproduct basis" introduced in Ref.~\cite{majrue}.\\
From (\ref{poinca}) one sees that the on-shell relation must be of the form~\cite{bob}
\begin{equation}
\mathcal{C} = E^2 - p^2 + \ell E p^2 ~.
\label{equation}
\end{equation}

Following again Ref.~\cite{bob} we 
derive the worldlines  within a Hamiltonian setup.
One starts  introducing canonical momenta conjugate
to the coordinates  $x$ and $t$:
\begin{eqnarray}
&& \{ p , x \} = -1~,~~ \{ E , t \} = 1~, \label{canonicalstandard}\\
&& \{ p , t \} = 0~,~~ \{ E , x \} = 0~. \nonumber
\end{eqnarray}
The on-shell relation $\mathcal{C}$ 
then will play the role of Hamiltonian.
Hamilton's equations give the conservation of $p$ and $E$ along the worldlines
\begin{equation}
\dot{p} = \frac{\partial \mathcal{C}}{\partial x} = 0 ~,
 ~~~ \dot E = - \frac{\partial \mathcal{C}}{\partial t} = 0 ~,
\end{equation}
where $\dot f \equiv \partial f/ \partial \tau$ and $\tau$ is an auxiliary worldline parameter.

The worldlines can then be obtained observing that
\begin{equation}
\begin{split}
&\dot{x} = - \frac{\partial \mathcal{C}}{\partial p} ~ \Rightarrow ~ x (\tau)
= x^{(0)} + ( 2 p
- 2 \ell \gamma_2 E p) \tau \\
&\dot t = \frac{\partial \mathcal{C}}{\partial E} ~ \Rightarrow ~ t(\tau) = t^{(0)} \!+\!\!\left(2 E + \ell p^2 \right) \tau~.
\end{split} \label{HamiltonEquations}
\end{equation}
Eliminating the parameter $\tau$  and imposing the Hamiltonian constraint $\mathcal{C} = 0$ (massless case) one finds that
\begin{equation}
x = x^{(0)} + \frac{p}{\sqrt{p^2+m^2}}  (t-t^{(0)}) - \ell p  (t-t^{(0)}) ~.
\label{fullworldlines}
\end{equation}
which reproduces (\ref{velocityDSR}) in the limit $m\rightarrow 0$.

The compatibility between boost transformations and form of the on-shell relation is encoded
in the requirement that the boost charge is conserved
\begin{equation}
\dot{\mathcal{N}} = \{ \mathcal{C} ,  \mathcal{N} \} = \frac{\partial  \mathcal{C}}{ \partial E}\frac{\partial  \mathcal{N} }{ \partial t} - \frac{\partial  \mathcal{C}}{ \partial p } \frac{\partial  \mathcal{N}}{ \partial x}  = 0 ~,
\end{equation}
This leads us to adopt the following representation of  boosts
\begin{equation}
\mathcal{N} = -t p + x E + \ell \, x \left( E^2 + \tfrac{1}{2} p^2 \right) ~.
\end{equation}
The DSR-relativistic covariance of this setup, ensured by construction, can also be verified
  by computing explicitly~\cite{bob} the action of an infinitesimal deformed boost with 
  rapidity vector $\xi$ 
\begin{eqnarray}
&&  p'= p - \xi E - \ell \xi \left( E^2 + \tfrac{1}{2} p^2 \right)
\label{ptra}\\ &&
 t' = t -  \xi x - 2 \ell \xi  E x
\label{ttra}\\ &&
 x' = x - t \xi + \ell \xi x p
\label{xtra}
\end{eqnarray}
Using these
one easily verifies that when Alice has the particle
on the worldline (\ref{fullworldlines}) Bob sees the particle on the worldline
\begin{equation}
 x' = x'^{(0)} + \frac{p}{\sqrt{m^2 + p'^2}} (t'-t'^{(0)}) - \ell \, p'(t'-t'^{(0)}) ~,
 \nonumber
\end{equation}
consistently with the (DSR-)relativistic nature of the framework.

Our next task is to summarize the evidence, first noticed indeed in Ref.~\cite{bob},
 that DSR-relativistic properties of the type illustrated by this
 phase-space construction may lead to the novel notion of a ``relativity of spacetime locality".\\
 Relativistically, as stressed in Ref.~\cite{bob}, the absoluteness of locality
 is codified in the fact that when one observer established that two events coincide
 then all other observers agree that the two events coincide.\\
 It had been known for several years~\cite{gacIJMPdsrREV,unruh,grilloSTDSR,sabinePRL}
 that a revision of the notion of locality is necessary in DSR-relativistic frameworks
 with momentum-dependent speed of massless particles. But a logically consistent picture
 did not arise until the study of Ref.~\cite{bob} showed that the needed revision
 of locality does not take the form of a total breakdown of locality, but rather
 just the weakening of locality that intervenes in going from an absolute-locality picture
 to a picture in which locality is a relativistic notion, assessed relatively to a
 given observer.\\
 We shall now summarize some of the results 
 of   Ref.~\cite{bob} which characterize this novel notion. but before we do that
 let us remind our readers of the analogy they should notice with the emergence of 
 relative simultaneity in special relativity. In Galilean relativity all velocities are relative
 and time is absolute. And it is well establish that the fact that Special Relativity
 introduce an absolute velocity scale $c$ came at the cost of rendering simultaneity relative.
 It is therefore not surprising if a DSR-relativistic framework, now endowing also to the
 momentums scale $|\ell|^{-1}$ with the role of absolute scale then must pay the price
 of rendering something relative, and this is where the relativity of locality is relevant.
 
\begin{figure}[H]
\begin{center}
\includegraphics[width=0.45\textwidth]{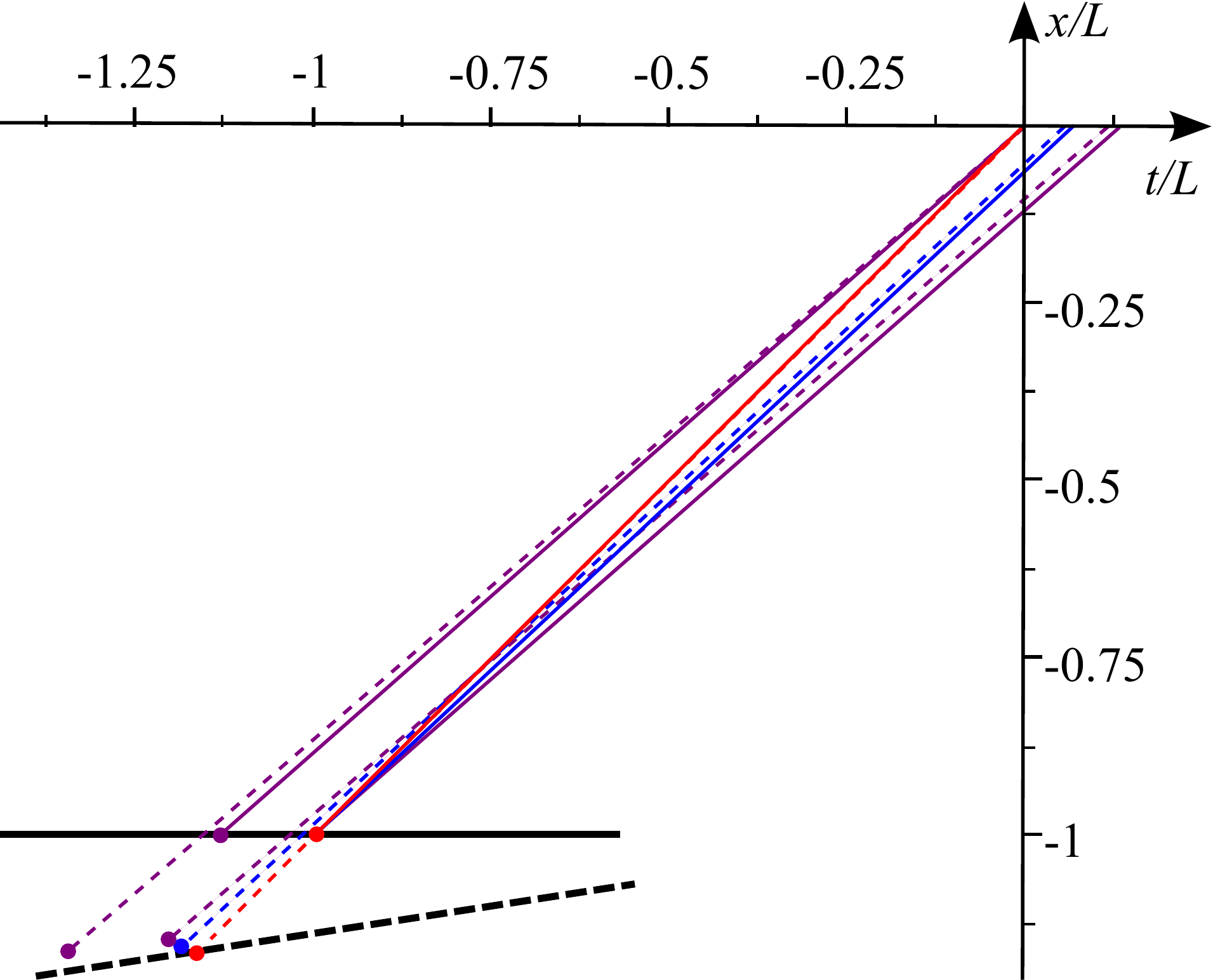}
\end{center}
\caption{A case with two ultrasoft (red) worldlines, still transforming as if boosts were undeformed, one with  ultrahard momentum $p_{v} = 0.1/\ell$ (violet), and one with momentum $p_{b}=p_{v}/2$ (blue).
According to Alice (whose lines are solid, while boosted Bob has dashed lines)
three of the worldlines give
a distant coincidence of events,
while two of the worldlines cross in the origin.}
  \label{fig1}
\end{figure}

 A quick and efficient characterization of the relativity of spacetime locality
 is obtained using the derivation of worldlines discussed above to obtain
 the results shown in Fig.~\ref{fig1}.

In figure we simply characterize ``events" as crossings of worldlines
(or crossing or a particle worldline with the worldline of an observer).
As shown by two of the worldlines in Fig.~\ref{fig1}, when an observer Alice is local
to a coincidence of
events ({\it i.e.} there are two events both occurring
in the origin of her coordinate system)
all observers that are purely boosted with respect to Alice, and therefore share
her origin of the coordinate system, will also describe those two events
as coincident. So in our DSR
framework one finds that locality,
a coincidence of events,
preserves its objectivity if assessed by a nearby observer.

The element of nonlocality that is actually produced by
DSR-deformed boosts is seen by focusing on the ``burst" of three photon worldlines
also shown in Fig.~\ref{fig1},
whose crossings establish a coincidence of events for Alice far from her origin,
an aspect of locality encoded in a ``distant coincidence of events".
The objectivity of such distant coincidences of events is partly spoiled by the
DSR deformation: as shown in Fig.~\ref{fig1}
the coincidence is only approximately present
in the coordinates of an observer boosted with respect to Alice.
We should however stress that even the distant coincidence is objective
 up to a very good approximation if indeed $|\ell|^{-1}$ is of the order of the
 Planck scale (corresponding to the Planck length $\sim 10^{-35}m$). 
 On terrestrial scales one might imagine hypothetically
to observe a certain particle decay with two  laboratories,
with a large relative boost of, say, $\xi \sim 10^{-5}$,
with idealized absolute accuracy in tracking   back
to the decay
region  the worldlines of two particles that are the decay products.
As one easily checks from (\ref{ttra})-(\ref{xtra}) the
 peculiar sort of departures from absolute locality that is codified
 in the ``relativity of spacetime locality" which we just described
 has magnitude governed by $\xi \ell L p$.
Therefore even if the distance $L$ between decay region and observers
if of, say,  $10^3 m$ and the decay products have momenta of, say, $100 GeV$,
one ends up with an apparent nonlocality of the decay region which is of
only $\sim 10^{-18}m$.

Going beyond terrestrial scales another context where estimates might be interesting
is the one of a typical observation of a gamma-ray burst, with particles of $GeV$
momenta that travel for, say, $10^{17}s$ before reaching our telescopes. In such a context,
for two satellite-telescopes
with a relative boost of, say, $\xi \sim 10^{-4}$,
the loss of coincidence of events at the source is of order $\sim 1Km$,
well below the sharpness we are able to attribute to the location of a gamma-ray burst.

We should also stress that actually in such a DSR
framework
two relatively boosted observers should not dwell about distant coincidences, but rather
express all observables in terms of local measurements (which is anyway
what should  be done in a relativistic theory). For example,
for the burst of three
photons shown in Fig.~\ref{fig1} the momentum dependence of the speed of photons
is objectively manifest (manifest both for Alice and Bob)
in the linear correlation between arrival times and momentum of the photons
(which we highlight in a panel contained in Fig.~\ref{fig1}).

Amusingly it appears
that coincident events were viewed as cumbersome
 already by Einstein, as shown by
a footnote in the famous 1905 paper~\cite{einstein}:
\vskip -0.2 cm
\begin{quote}
{\small  ``{\it We shall not discuss here the imprecision inherent in the concept of
simultaneity of two events taking place at (approximately) the same
location, which can be removed only by abstraction.}"}
\end{quote}

\section{$\kappa$-Poincar\'e phase-space construction with ``$\kappa$-Minkowski coordinates"}

\subsection{A $\kappa$-Minkowski {\underline{coordinate}} velocity}
Our next task is to summarize the results of Ref.~\cite{k-bob}
with a $\kappa$-Poincar\'e phase-space construction
using ``$\kappa$-Minkowski coordinates", meaning that
we shall assume  ``$\kappa$-Minkowski Poisson brackets"
for the spacetime coordinates
 \begin{equation}
\left\{ x,t\right\} =-\ell x ~ .
\label{kappadef}
\end{equation}
We label these as ``$\kappa$-Minkowski Poisson brackets"
because they mimic the noncommutativity $[ {\hat{x}} ,{\hat{t}} ] =-\ell {\hat{x}}$
of the $\kappa$-Minkowski spacetime~\cite{majrue,lukieANNALS}.

Notice that one goes from the standard coordinates of the previous section to 
the ``$\kappa$-Minkowski coordinates" of (\ref{kappadef}) by using the
redefinition $x \rightarrow x$, $t \rightarrow  t + \ell xP $,
which was already considered for other reasons in Ref.~\cite{leeINERTIALlimit}.
And we stress that the
 physical aspects of this sort of relativistic theories
are all codified in the readout of
clocks local to the emission of particles
and (appropriately synchronized) clocks local to the detection of particles.
So the redefinition $x \rightarrow x$, $t \rightarrow  t + \ell xP$,
must be armless: the redefinition is moot at $x=0$ so it has no
effect for the times an observer assigns to emission or detection
events that she witnesses as ``nearby observer" 
(in her origin with $x=0$). We must expect that the results in this section will confirm
the results of the previous section, although in a peculiar disguise.

The description of space and time translations is obtained from (\ref{canonicalstandard})
taking only into account the redefinition $x \rightarrow x$, $t \rightarrow  t + \ell xP$:
\begin{gather}
\left\{ E ,t\right\} =1,\qquad\left\{ E ,x\right\} =0\ ,\label{timetrasl}
\\ \left\{ p,t\right\} =\ell p,\qquad\left\{ p,x\right\} =-1~.
\label{spacetrasl} \end{gather}
Notice that the difference between these translations and the ones of (\ref{canonicalstandard}) all resides
in $\{ p,t \} =\ell p$. But this must be taken into account consistently throughout~\cite{k-bob}
the analysis. We shall see that it brings about changes for the derivation of the ``coordinate
velocity", and, as observed in Ref.~\cite{k-bob}, 
 it also changes the form of the map between observers in such a way that the ``coordinate velocity"
 does not coincide with the physical velocity.

In comparing the content of this section and of the previous one it is necessary
to take into account of this difference of translation generators, whereas
also in this section we use the same description of boost transformations of Eq.~(\ref{poinca})
 (and of course we still maintain the commutativity of translations):
\begin{equation}
\left\{ E ,p\right\} =0 \, ,~\,
 \left\{ \mathcal{N}, E \right\} = p \, ,~\,
 \left\{ \mathcal{N},p \right\} =E + \ell E^{2} + \frac{\ell}{2}p^{2}~,
\end{equation}
which of course also preserves the form of the on-shell relation
\begin{equation}
m^2=E^{2}-p^{2}+\ell E p^{2} ~.
\end{equation}
It is easy to check that (\ref{kappadef}),(\ref{timetrasl}),(\ref{spacetrasl}),(\ref{poinca})
satisfy all Jacobi identities.

Let us use again the on-shell relation
as Hamiltonian
of evolution of the observables on the worldline of a particle
in terms of the worldline parameter $\tau$.
Again Hamilton's equations give the conservation of $p$ and $E$ along the worldlines
\begin{equation}
\dot{p} = \frac{\partial \mathcal{C}}{\partial x} = 0 ~,
 ~~~ \dot{E} = - \frac{\partial \mathcal{C}}{\partial t} = 0 ~.\nonumber
\end{equation}

One then takes into account (\ref{spacetrasl}) in the derivation
of the equations of motion:
\begin{gather}
\dot{t}=\left\{ \mathcal{C},t\right\}
=\frac{\partial\mathcal{C}}{\partial E}\left\{ E ,t\right\}
+\frac{\partial\mathcal{C}}{\partial p}\left\{ p,t\right\} = 2E
-\ell p^{2} \ ,\nonumber \\
\dot{x}=\left\{ \mathcal{C},x\right\}
=\frac{\partial\mathcal{C}}{\partial E}\left\{ E ,x\right\}
+\frac{\partial\mathcal{C}}{\partial p}\left\{ p,x\right\} = 2p-2\ell E p
~. \nonumber \end{gather}

 This evidently implies
$t\left(\tau\right)=t_{0}+\left(2E-\ell p^{2}\right)\tau$ and
$x\left(\tau\right)=x_{0}+\left(2p-2\ell Ep\right)\tau$,
from which, eliminating the parameter $\tau$ and imposing the Hamiltonian
constraint $\mathcal{C}=m^{2}$, one finds
\begin{equation}
x\left(p,x_0,t_0;t\right) \!
= \! x_{0} \! + \! \left(\frac{p}{\sqrt{p^{2}\! +\! m^{2}}} \!
- \! \ell p\left(1 \!
- \! \frac{p^{2}}{p^{2} \! + \! m^{2}}\right)\right)\left(t \! - \! t_{0}\right)
\nonumber \end{equation}
In particular, for massless particles these worldlines give
a momentum-independent coordinate velocity:
\begin{equation}
x \left(p,x_0,t_0;t\right)=x_{0}+(t-t_{0})p/|p|\ .\label{momindep}\end{equation}

This momentum-independent coordinate velocity had been derived
in several previous studies (see, {\it e.g.},
Refs.~\cite{jurekvelISOne,lukieVEL,kosinskiVEL,ghoshVELisONE,mignemiVEL}),
and these studies, which were unaware of the possibility of relative locality,
emphasized a possible contrast between this result of a possible analysis
of classical phase spaces inspired by $\kappa$-Minkowski and arguments
based on the quantum properties of $\kappa$-Minkowski spacetime, which had
found~\cite{gacmaj,gacMandaniciDANDREA} the momentum dependence of the speed
of massless particles.

\subsection{What about Bob?}

Awareness of the possibility of a relativity of spacetime locality,
in the sense of Ref.~\cite{bob} (here summarized in the previous section)
immediately changes one's perspective on the momentum independence
of the coordinate velocity of massless particles found in Eq.~(\ref{momindep}).\\
Let us expose this issue by following the reasoning of Ref.~\cite{k-bob}.\\
It suffices to
 formalize
a simultaneous emission occurring in the origin of an observer Alice.
This will be described by Alice in terms of two worldlines, a massless particle
 with momentum $p_1$ ($>0$)
and a massless particle
with momentum $p_2$ ($>0$), which actually coincide because of the momentum independence
of the coordinate velocity:
\begin{gather}
x^{A}_{p_1}(t^{A})=t^{A} ~,~~~
x^{A}_{p_2}(t^{A})=t^{A}~. \label{alicep}
\end{gather}
It is useful to focus on the case of $p_1$ and $p_2$ such that $|p_1| \ll |p_2|$,
and $|\ell p_1| \simeq 0$
(the particle with momentum $p_1$ is soft enough that it behaves as if $\ell = 0$) 
while $|\ell p_2| \neq 0$, in the sense
 that for the hard particle the effects of $\ell$-deformation  are not negligible.

A central role in our analysis is played by
the translation transformations codified in (\ref{timetrasl}),(\ref{spacetrasl}).
These allow us to establish how the assignment of coordinates
on points of a worldline differs between two observers connected
by a generic translation ${\cal T}_{a_t , a_x}$, with component $a_{t}$
along the $t$ axis and component $a_{x}$ along the $x$
axis~\cite{arzkowaRelLoc}
\begin{gather}
x'=x-a_{t}\left\{ E ,x\right\} +a_{x}\left\{ p,x\right\} \ ,\nonumber \\
t'=t-a_{t}\left\{ E ,t\right\} +a_{x}\left\{ p,t\right\} \ .\nonumber
\end{gather}

Using these we can look at the two Alice worldlines, given in (\ref{alicep}),
from the perspective
of a second observer, Bob, at rest with respect to Alice at distance $a$ from Alice
(Bob = ${\cal T}_{a , a} \triangleright$ Alice),
local to a detector that the two particles eventually reach.
Of course, since we have seen that the coordinate velocity is momentum independent,
according to Alice's coordinates the two particles reach Bob simultaneously.\\
But can this distant coincidence of events be trusted?\\
The two events which according to the coordinates of distant observer Alice are coincident
 are the crossing of Bob's worldline with the worldline of the particle
with momentum $p_1$ and the
crossing of Bob's worldline with the worldline of the particle
with momentum $p_2$. This is a coincidence of events distant from observer Alice,
which in presence of the relativity of spacetime locality cannot be trusted.\\
To clarify the situation we should look at the two worldlines from the perspective
of Bob, the observer who is local to the detection of the particles.

Evidently these Bob worldlines are obtained from Alice worldlines using
the translation transformation codified in (\ref{timetrasl}),(\ref{spacetrasl}).
Acting on a generic Alice worldline $x^A(p^A,x^A_0,t^A_0;t^A)$
this gives a Bob worldline $x^B(p^B,x^B_0,t^B_0;t^B)$ as follows:
\begin{gather}
p^B = p^A - a \left\{ E ,p^A\right\} +a \left\{ p,p^A\right\}
= p^A
 \ ,\nonumber \\
x_0^B = x_0^A - a \left\{ E ,x_0^A\right\} +a \left\{ p,x_0^A\right\}
= x_0^A-a
 \ ,\nonumber \\
t_0^B=t_0^A - a \left\{ E ,t_0^A\right\} +a \left\{ p,t_0^A\right\}
=t_0^A-a +\ell a p
  \ .\nonumber \end{gather}
  And specifically for the two worldlines of our interest, given for Alice in
  (\ref{alicep}),
  one then finds
\begin{gather}
x^{B}_{p_1}(t^{B})=t^{B}-\ell a p_1 \simeq t^{B}
 \ ,\nonumber \\
x^{B}_{p_2}(t^{B})=t^{B}-\ell a p_2\ .\nonumber \end{gather}
We have found that,
because of the peculiarities of translational symmetries
of the $\kappa$-Minkowski quantum spacetime,
the two worldlines, which were coincident according to Alice, are
distinct worldlines for Bob.
According to Bob, who is at the detector, the two particles reach the detector
at different times: $t^B \simeq 0$ for the soft particle and
$t^{B} = \ell a p_2$ for the hard particle. And these are the two
massless particles
which, according to the observer Alice who is at the emitter, were emitted simultaneously.


\begin{figure}[H]
\begin{center}
\includegraphics[width=0.33 \textwidth]{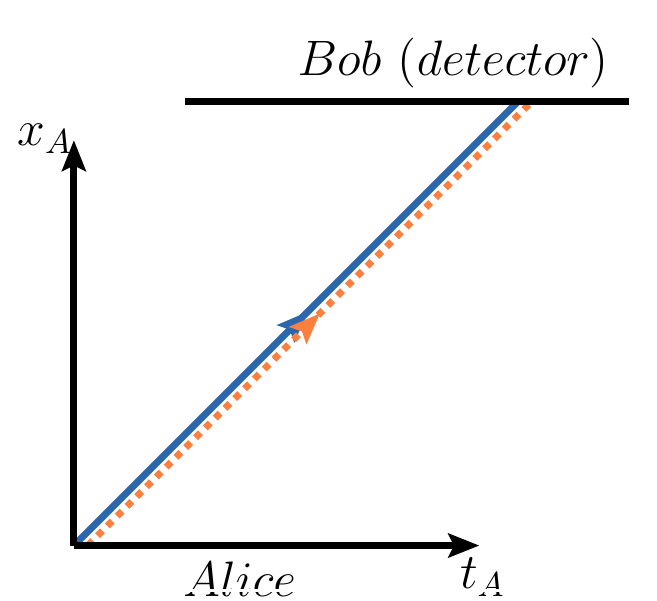}
\includegraphics[width=0.37 \textwidth]{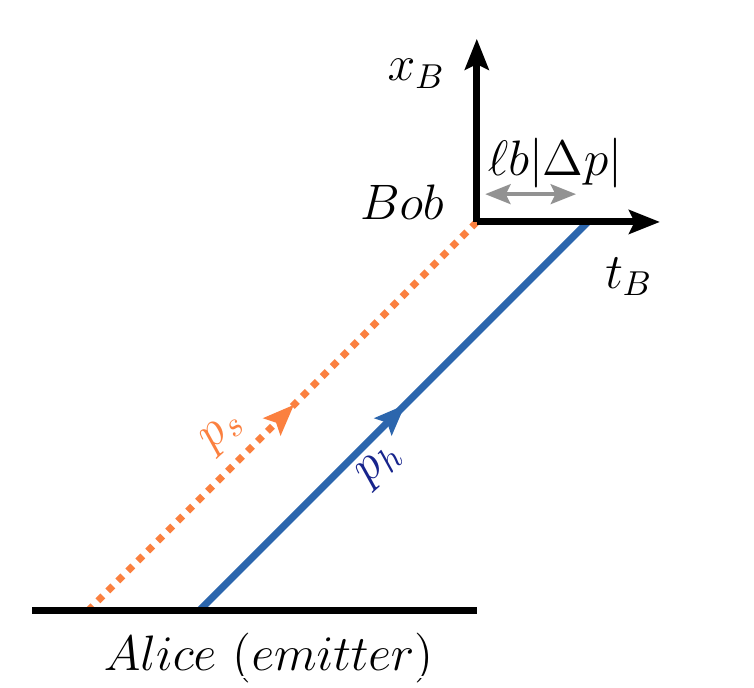}
\caption{\label{alicekappabob} Two simultaneously-emitted massless particles
of different momentum in $\kappa$-Minkowski
 are detected at different times.
 The figure shows
 how the simultaneous emission of two such particles and their
 non-simultaneous detection
 who is at the emitter,
 and in the ``$\kappa$-Minkowski coordinates" of observer Bob (right panel),
 who is at the detector.}
\end{center}
\end{figure}

\noindent
Reassuringly this result  of Ref.~\cite{k-bob} for momentum dependence of time of detection
of simultaneously-emitted massless particles,
$$\Delta t = \ell \Delta p~,$$
is in perfect agreement with the result already obtained in Ref.~\cite{bob} (here summarized in the
previous section) and with the findings of Refs.~\cite{gacmaj,gacMandaniciDANDREA}, which,
using completely independent arguments, had found a result for the physical velocity
in $\kappa$-Minkowski that is in perfect agreement (including signs and numerical factors)
with these quantitative prediction
for the momentum dependence of time of detection
of simultaneously-emitted massless particles.

It is perhaps worth highlighting the soundness of the operative
procedure by which we have determined this correlation between
momentum of simultaneously-emitted particles
and times of detection.
Our procedure
rests safely on the robust shoulders of the procedure for
determining the physical velocity measured by inertial observers in classical Minkowski
spacetime, and this connection is allowed by the fact that the properties
of infrared massless particles (properties of massless particles in the infrared limit)
are unaffected by the $\kappa$-Poincar\'e deformation.
The distant synchronization of the clocks
on emitter/Alice and on detector/Bob is evidently a special-relativistic synchronization,
relying on exchanges of infrared massless particles.
And we implicitly assumed that
the relative rest of Alice and Bob is established
by exchanges of infrared massless particles, so that indeed it can borrow
from the special-relativistic operative definition of
inertial observers in relative rest.
By construction the distance $a$ between Alice and Bob
is also defined operatively just like in special relativity, with the only
peculiarity that Alice and Bob should determine it by exchanging
infrared massless particles.
This setup guarantees that infrared massless particles
are timed and observed
in $\kappa$-Minkowski exactly as in classical Minkowski spacetime.
The new element of the $\kappa$-Minkowski relativistic theory,
concerning the  ``hard" (``high-momentum") massless particles,
then also acquires a sound operative definition by our procedure centered
on the simultaneous emission (with simultaneity prudently established
by the local observer/emitter Alice) of an infrared and a hard massless particle,
then comparing the arrival times at observer/detector Bob (times of arrival
 prudently established according to the local observer, indeed Bob).

In the idealized setting of a sharply flat spacetime our procedure
is applicable for any arbitrarily high value of
the distance $a$. But of course for most realistic applications one will be interested
 in contexts where sharp flatness
of spacetime cannot a priori be assumed, and evidently in such more general cases
the limit $a \rightarrow 0$ of our procedure should be relied upon.
Note however that the characterization of observations (local) and inferences (distant)
given by the coordinates of Alice and Bob, which we summarized in Fig.~\ref{alicekappabob},
evidently remains valid even for small values of $a$: no matter how close Alice and Bob
are, one still has that in Alice's  coordinates the detections at Bob appear
to be simultaneous (while Bob, local to the detections, establishes that they are
not simultaneous) and that
in Bob's  coordinates the emission at Alice appears
to be not simultaneous
(while Alice, local to the emissions, establishes that they are
simultaneous).

\section{$\kappa$-Poincar\'e inspired relative-locality momentum space}

\subsection{The ``relative-locality framework"}
In the previous two sections we summarized the results that establish the
momentum dependence of the speed of massless particles in certain $\kappa$-Poincar\'e inspired
 phase space, and we saw already a few aspects of the relativity of spacetime locality. All this
 however was confined to formalizations that can only accommodate free particles, which
 evidently is not fully satisfactory.
 On the other hand for nearly 15 years it was unknown how to formulate
 interactions on a $\kappa$-Poincar\'e phase space. One of the major obstructions
 for progress in this direction 
 came from the fact that several arguments (recently reviewed in Ref.~\cite{anatomy})
 suggest that the law of composition of momenta on a $\kappa$-Poincar\'e phase space
 of the type we are studying should be of the type
\begin{equation}
 (p \oplus q)_\mu = p_\mu +q_\mu -\ell \delta^j_\mu p_0 q_j
 \label{mrcomposition}
\end{equation}
and it was unclear how to implement this in a consistent manner.

The answer to this puzzles came only recently and in a round-about manner.
First came the proposal, in Refs.~\cite{prl,grf2nd}, of the ``relative locality framework"
which can accommodate a vast class of proposals for modifications of the on-shell relation
and modifications of the law of composition of momenta, while providing indeed
a coherent description of interactions among particles. Then came the realization~\cite{anatomy}
(also see Ref.~\cite{flagiuKAPPAPRL})
that a particular application of the relative-locality framework would
allow to address the longstanding issues for the description of interactions
on $\kappa$-Poincar\'e inspired
 phase spaces.
 
The relative-locality framework has an extremely rich spectrum of properties and
possible applications, for which we direct our readers to 
Refs.~\cite{prl,grf2nd,leelaurentGRB,soccerball,flagiuKAPPAPRL,anatomy,flajosePRL,goldenrule}.
We shall here focus on summarizing its properties which are relevant for extending
the scopes of the topics we discussed in the previous two sections to the case
of interacting particles on a $\kappa$-Poincar\'e inspired
 phase space.

The relative-locality framework is centered on the geometry of momentum space.
It is a framework suitable for the study of possibly deviations from the predictions
of the standard picture of momentum space, which assumes for it the 
flat geometry of a Minkowski space.
On momentum space one introduces~\cite{prl,grf2nd}
a metric and an affine connection.
The momentum-space metric $ds^{2}= g^{\mu\nu}(p) {\mathrm d}p_{\mu}{\mathrm d}p_{\nu}$ 
characterized the
energy-momentum on-shell relation
$$m^{2} = D^{2}(p)$$
where $D(p)$ is the distance of the point $p_{\mu}$ from the origin $p_{\mu}=0$.\\
The affine connection  characterized  the law of composition of momenta
$$\left(p \oplus q\right)_\mu \simeq p_\mu + q_\mu -
\ell \Gamma_\mu^{\alpha\beta}\, p_\alpha\,
q_\beta + \cdots$$
where the right-hand side assumes momenta are small with respect to $|\ell|^{-1}$
 and $\Gamma_{\mu}^{\alpha\beta}$
 are the ($\ell$-rescaled) connection coefficients on momentum space
 evaluated at $p_\mu = 0$.

Already in Ref.~\cite{prl} an action principle was given that could govern
a single-interaction process on the relative-locality momentum space
\begin{equation*}
 {\cal S} = \int ds \left( x^\mu_J \dot{p}_\mu^J -{\cal N}_J {\cal C}^J (p) \right) - \xi^\mu \cal{K}_\mu
\end{equation*}
The bulk part of this action ends up characterizing~\cite{prl} the
propagation of the particles, with the Lagrange multipliers
$\mathcal{N}_J$ 
enforcing the on-shell relations ${\cal C}^J[k]= D^2(k)-m_J^2$,
and $D^2(k)$ in turn derived from the metric on momentum space
as the distance of $k_\mu$ from the origin
of momentum space.
The form of the boundary term $\xi^{\mu}\mathcal{K}_\mu(s_{0})$
is such that~\cite{prl} the Lagrange multipliers $\xi^{\mu}$
enforce the condition $\mathcal{K}_\mu(s_{0}) = 0$,
so that by taking for $\mathcal{K}_\mu$ a suitable composition
of the momenta $k_\mu$, $p_\mu$, $q_\mu$ the boundary terms enforces
a law of conservation of momentum at the interaction.

For our purposes here, and for the work reported in Ref.~\cite{anatomy} which
we here summarize, it is important to notice that the on-shell relation
we used in the previous two sections for our discussions of a $\kappa$-Poincar\'e phase space
can be formulated in this framework by adopting a de Sitter metric on 
momentum space~\cite{flagiuKAPPAPRL,anatomy}.
And the law of composition of momenta (\ref{mrcomposition}) which is of interest
for the development of this phase-construction can be faithfully described
as a legitimate (but non-metric and torsionful) choice of affine connection on a
de-Sitter momentum space~\cite{flagiuKAPPAPRL,anatomy}. So one can rely on
the powerful machinery of the relative-locality framework for introducing
interactions on a $\kappa$-Poincar\'e phase space.

In the rest of this section we shall summarize the findings Ref.~\cite{anatomy}, which
were based on this strategy of description of interactions 
on a $\kappa$-Poincar\'e phase space.

In this section we adopt the conventions of Ref.~\cite{anatomy}, 
which also allow an easier comparison to other results
on the relative-locality framework. With respect to the conventions used in the previous
two sections this consists in mapping
\begin{equation}
E \rightarrow p_0 \ , \qquad p \rightarrow - p_1 \ , \qquad t \rightarrow x^0 \ , \qquad x \rightarrow x^1 \ ,
\label{maptoanatomy}
\end{equation}
together with a sign change for
Poisson brackets
\begin{equation}
\{ \ , \ \} \rightarrow  \ - \, \{ \ , \ \} \ .
\end{equation}
 We also often use in this section
 the compact notation for which $p_\mu \equiv (p_0 , p_1) $ and $p^\mu \equiv (p^0 , p^1)$, 
 and then the map (\ref{maptoanatomy}) turns to be convenient\footnote{One can also think that if $(E,p,t,x) \equiv (p^0,p^1,x^0,t^1)$, with up indexes, the map (\ref{maptoanatomy}) is equivalent to lower the indexes of the momenta with a Lorentzian metric $\{-1,1 \}$.}
for expressing the action of momenta,
when they act as generators of translations, in compact form, as in
\begin{equation}
x^\mu \rightarrow x^\mu + b^\nu \{ p_\nu , x^\mu \} \ .
\end{equation}

\subsection{Causally connected interactions and translations
generated by total momentum}\label{main}
The primary objective of the study reported in Ref.~\cite{anatomy} was to
provide a prescription for achieving translational invariance of the description,
within the relative-locality framework, of cases with causally-connected interactions.
It was shown in  Ref.~\cite{anatomy} that there are alternative ways to codify momentum-conservation
laws in boundary conditions, but the requirement of translational invariance
of the description of causally-connected interactions
severely restricts the number of options that are available.
Successfully finding ways to satisfy these restrictions was the main
achievement reported in Ref.~\cite{anatomy}, but we shall here not focus on that issue.
Rather we take the translationally-invariant description devised in Ref.~\cite{anatomy}
as an established starting point, and summarize here instead the parts of 
Ref.~\cite{anatomy} which are relevant for the issues here of interest concerning
the speed of massless particles on $\kappa$-Poincar\'e phase spaces and
the relativity of spacetime locality.

\begin{figure}[h!]
\begin{center}
\includegraphics[width=0.66 \textwidth]{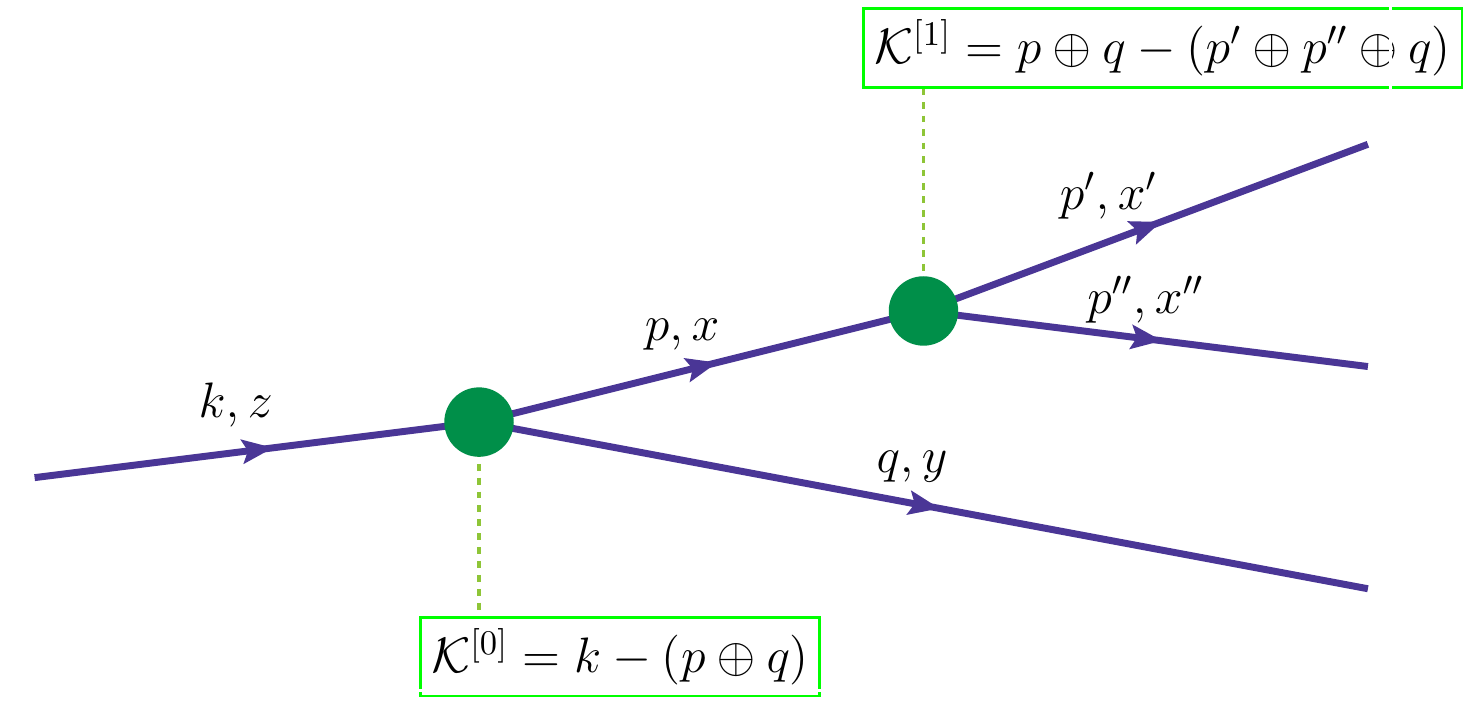}
\caption{\label{fixin2vertic} The case of causally-connected interactions analyzed in this subsection.}
\end{center}
\end{figure}

For this we only need the part of 
Ref.~\cite{anatomy} which concerned a single pair of causally-connected interactions,
as in the case we here show in Fig.~\ref{fixin2vertic}.

Following Ref.~\cite{anatomy} we describe the case shown in Fig.~\ref{fixin2vertic} through the action
\begin{equation}
\begin{split}
\mathcal{S}^{\kappa (2)} = & \int_{-\infty}^{s_{0}}ds\left(z^{\mu}\dot{k}_{\mu}-\ell z^{1}k_{1}\dot{k}_{0}+\mathcal{N}_k\mathcal{C}_\kappa\left[k\right]\right)+
\int_{s_{0}}^{s_{1}}ds\left({x}^{\mu}\dot{p}_{\mu}-\ell x^{1}p_{1}\dot{p}_{0}+\mathcal{N}_p\mathcal{C}_\kappa\left[p\right]\right)\\
   +\!&\int_{s_{1}}^{+\infty}\!\!
ds\left({x'}^{\mu}\dot{p'}_{\mu}-\ell{x'}^{1}p'_{1}\dot{p'}_{0}+\mathcal{N}_{p'}\mathcal{C}_\kappa\left[p'\right]\right)+\int_{s_{1}}^{+\infty}\!\!ds\left(x''^{\mu}\dot{p''}_{\mu}-\ell x''^{1}p''_{1}\dot{p''}_{0}+\mathcal{N}_{p''}\mathcal{C}_\kappa\left[p''\right]\right)\\
 &+\int_{s_{0}}^{+\infty}ds\left(y^{\mu}\dot{q}_{\mu}-\ell y^{1}q_{1}\dot{q}_{0}+\mathcal{N}_q\mathcal{C}_\kappa\left[q\right]\right) -\xi_{[0]}^{\mu}\mathcal{K}_{\mu}^{[0]}(s_{0})-\xi_{[1]}^{\mu}\mathcal{K}_{\mu}^{[1]}(s_{1})\ ,
 \label{actionk3+3full}
\end{split}
\end{equation}
 where the boundary terms at endpoints of worldlines a given, following
 the prescription of Ref.~\cite{anatomy}, in terms of
\begin{equation*}
\begin{split} &
{\cal K}^{[0]} = k-(p\oplus q) \\ &
{\cal K}^{[1]} = (p\oplus q) - ( p'\oplus p'' \oplus q)  \ ,
\end{split}
\end{equation*}
{\it i.e.}
\begin{gather}
{\cal K}^{[0]}_\mu = k_\mu -(p\oplus q)_\mu  = k_\mu - p_\mu -q_\mu -\ell \delta _\mu ^1 p_0 q_1 \ ,\nonumber\\
{\cal K}^{[1]}_\mu = (p\oplus q)_\mu - ( p'\oplus p'' \oplus q)_\mu = p_\mu  - p_\mu'- p_\mu'' - \ell \delta_\mu^1 \left(-p_0 q_1 + p_0' p_1'' + p_0' q_1 + p_0'' q_1 \right) \ .
\label{K[1]}
\end{gather}

This formulation ensures the availability
of a relativistic description of distant observers, {\it i.e.}
the invariance under translation transformations.\\
In order to see that this is the case
it suffices to note down the
 equations
 of motion (and constraints)
 that follow from the action $\mathcal{S}^{\kappa (2)}$:
\begin{gather}
\dot p_\mu =0~,~~\dot q_\mu =0~,~~\dot k_\mu =0~,~~\dot p'_\mu =0~,~~\dot p''_\mu =0\ ,\nonumber\\
\mathcal{C}_{\kappa}[p]=0~,~~\mathcal{C}_{\kappa}[q]=0~,~~\mathcal{C}_{\kappa}[k]=0~,~~\mathcal{C}_{\kappa}[p']=0~,~~\mathcal{C}_{\kappa}[p'']=0\label{eqmotion3+3kmomentum}\ ,\\
\mathcal{K}_\mu^{[0]}(s_{0})=0~,~~\mathcal{K}_\mu^{[1]}(s_{1})=0\nonumber\ ,
\end{gather}
\begin{gather}
\dot x^\mu = \mathcal{N}_p \left(\frac{\delta \mathcal{C}_{\kappa}[p]}{\delta p_\mu} +\ell \delta^\mu_0 \frac{\delta \mathcal{C}_{\kappa}[p]}{\delta p_1} p_1\right)= \delta^{\mu}_{0}\mathcal{N}_p \left( 2 p_{0}-\ell p_{1}^{2}\right)-2\delta^{\mu}_{1}\mathcal{N}_p \left(  p_{1}-\ell p_{0} p_{1}\right)\nonumber \ ,\\
\dot y^\mu = \mathcal{N}_q \left(\frac{\delta \mathcal{C}_{\kappa}[q]}{\delta q_\mu} +\ell \delta^\mu_0 \frac{\delta \mathcal{C}_{\kappa}[q]}{\delta q_1} q_1\right)= \delta^{\mu}_{0}\mathcal{N}_q \left( 2 q_{0}-\ell q_{1}^{2}\right)-2\delta^{\mu}_{1}\mathcal{N}_q \left(  q_{1}-\ell q_{0} q_{1}\right)\nonumber \ ,\\
\dot z^\mu = \mathcal{N}_k \left(\frac{\delta \mathcal{C}_{\kappa}[k]}{\delta k_\mu} +\ell \delta^\mu_0 \frac{\delta \mathcal{C}_{\kappa}[k]}{\delta k_1} k_1\right)= \delta^{\mu}_{0}\mathcal{N}_k \left( 2 k_{0}-\ell k_{1}^{2}\right)-2\delta^{\mu}_{1}\mathcal{N}_k \left(  k_{1}-\ell k_{0} k_{1}\right)\nonumber \ ,\\
\dot x'^\mu =\mathcal{N}_{p'} \left(\frac{\delta \mathcal{C}_{\kappa}[p']}{\delta p'_\mu} +\ell \delta^\mu_0 \frac{\delta \mathcal{C}_{\kappa}[p']}{\delta p'_1} p'_1\right)=\delta^{\mu}_{0}\mathcal{N}_{p'} \left( 2 p_{0}'-\ell p_{1}'^{2}\right)-2\delta^{\mu}_{1}\mathcal{N}_{p'} \left(  p_{1}'-\ell p_{0}' p_{1}'\right)\ ,
\label{eqmotion3+3kspace}\\
\dot x''^\mu = \mathcal{N}_p'' \left(\frac{\delta \mathcal{C}_{\kappa}[p'']}{\delta p''_\mu} +\ell \delta^\mu_0 \frac{\delta \mathcal{C}_{\kappa}[k]}{\delta p''_1} p''_1\right)=\delta^{\mu}_{0}\mathcal{N}_{p''} \left( 2 p_{0}''-\ell p_{1}''^{2}\right)-2\delta^{\mu}_{1}\mathcal{N}_{p''} \left(  p_{1}''-\ell p_{0}'' p_{1}''\right)
~,\nonumber
\end{gather}
and the conditions at the $s=s_{0}$ and $s=s_{1}$ boundaries:
\begin{gather}
z^\mu(s_{0}) = \xi^\nu_{[0]} \left(\frac{\delta \mathcal{K}^{[0]}_\nu}{\delta k_\mu}+\ell \delta^\mu_0 \frac{\delta \mathcal{K}^{[0]}_\nu}{\delta k_1}k_1\right)=\xi^\mu_{[0]} +\ell \delta^{\mu}_{0} \xi^1_{[0]} k_{1}\nonumber\ ,\\
x^\mu(s_{0}) = -\xi^\nu_{[0]} \left(\frac{\delta \mathcal{K}^{[0]}_\nu}{\delta p_\mu}+\ell \delta^\mu_0 \frac{\delta \mathcal{K}^{[0]}_\nu}{\delta p_1}p_1\right)=\xi^\mu_{[0]} +\ell \delta^{\mu}_{0} \xi^1_{[0]} (p_{1}+q_{1})\ , \nonumber\\
x^\mu(s_{1}) = \xi^\nu_{[1]} \left(\frac{\delta \mathcal{K}^{[1]}_\nu}{\delta p_\mu}+\ell \delta^\mu_0 \frac{\delta \mathcal{K}^{[1]}_\nu}{\delta p_1}p_1\right)=\xi^\mu_{[1]} +\ell \delta^{\mu}_{0} \xi^1_{[1]} (p_{1}+q_{1})\nonumber\ ,\\
y^\mu(s_{0})= -\xi^\nu_{[0]} \left(\frac{\delta \mathcal{K}^{[0]}_\nu}{\delta q_\mu}+\ell \delta^\mu_0 \frac{\delta \mathcal{K}^{[0]}_\nu}{\delta q_1}q_1\right)=\xi^\mu_{[0]} +\ell \delta^{\mu}_{0} \xi^1_{[0]} q_{1}+\ell \delta^{\mu}_{1} \xi^1_{[0]} p_{0}
\label{boundaries3+3k}\ ,\\
x'^\mu(s_{1}) = -\xi^\nu_{[1]} \left(\frac{\delta \mathcal{K}^{[1]}_\nu}{\delta p'_\mu}+\ell \delta^\mu_0 \frac{\delta \mathcal{K}^{[1]}_\nu}{\delta p'_1}p'_1\right)=\xi^\mu_{[1]} +\ell \delta^{\mu}_{0} \xi^1_{[1]} (p_{1}'+p_{1}''+q_{1})\nonumber\ ,\\
x''^\mu(s_{1}) = -\xi^\nu_{[1]} \left(\frac{\delta \mathcal{K}^{[1]}_\nu}{\delta p''_\mu}+\ell \delta^\mu_0 \frac{\delta \mathcal{K}^{[1]}_\nu}{\delta p''_1}p''_1\right)=\xi^\mu_{[1]} +\ell \delta^{\mu}_{0} \xi^1_{[1]} (p_{1}''+q_{1})+\ell \delta^{\mu}_{1} \xi^1_{[1]} p_{0}'\nonumber\ .
\end{gather}
One then can easily check~\cite{anatomy} that the following translation
transformations, generated by the total momentum,
\vspace{-0.5cm}
\begin{equation}
\begin{split}
z_{B}^{0}(s)&=  z_{A}^{0}(s)+b^{\mu} \{ k_{\mu},z^{0}\}
= z_{A}^{0}(s)-b^{0}-\ell b^{1}k_{1}\ ,\\
z_{B}^{1}(s)&=z_{A}^{1}(s)+b^{\mu} \{ k_{\mu},z^{1}\}
={z}_{A}^{1}(s)-b^{1}\ ,\\
x_{B}^{0}(s)&=  x_{A}^{0}(s) + b^\mu
 \{ (p\oplus q)_\mu , x^0\}
=x_{A}^{0}(s) - b^{0}-\ell b^{1}(p_{1}+q_1)\ , \\
x_{B}^{1}(s)&=x_{A}^{1}(s)+b^{\mu}
 \{ (p\oplus q)_\mu , x^0\}
={x}_{A}^{1}(s)-b^{1}\ ,\\
y_{B}^{0}(s)&=  y_{A}^{0}(s)+b^{\mu}  \{ (p\oplus q)_\mu , y^0\}
=y_{A}^{0}(s)-b^{0}-\ell b^{1}q_1\ ,\\
y_{B}^{1}(s)&=y_{A}^{1}(s)+b^{\mu} \{ (p\oplus q)_\mu , y^1\}
={y}_{A}^{1}(s)-b^{1}-\ell b^{1}p_0\ ,\\
{x'}_{B}^{0}(s)&={x'}_{A}^{0}(s)+b^{\mu} \{ (p'\oplus p'' \oplus q)_\mu , x'^0\}
={x'}_{A}^{0}(s)-b^{0}-\ell b^{1}(p'_{1}+p''_{1}+q_1)\ , \\
{x'}_{B}^{1}(s)&={x'}_{A}^{1}(s)+b^{\mu} \{ (p'\oplus p'' \oplus q)_\mu , x'^1\}
={x'}_{A}^{1}(s)-b^{1}\ ,\\
{x''}_{B}^{0}(s)&={x''}_{A}^{0}(s)+b^{\mu}
 \{ (p' \oplus p'' \oplus q)_\mu , x''^0\}
={x''}_{A}^{0}(s)-b^{0}-\ell b^{1}(p''_{1}+q_1)\ , \\
{x''}_{B}^{1}(s)&={x''}_{A}^{1}(s)+b^{\mu}
 \{ (p' \oplus p'' \oplus q)_\mu , x''^1\}
={x''}_{A}^{1}(s)-b^{1}-\ell b^1 p'_0\ .
\label{translations3+3k}
\end{split}
\end{equation}
leave the equations of motion (\ref{eqmotion3+3kspace}) unchanged 
and leave the boundary conditions  (\ref{translations3+3k}) unchanged.

\subsection{Free-particle limit}\label{kappabobprl}

\begin{figure}[H]
\begin{center}
\includegraphics[width=0.72 \textwidth]{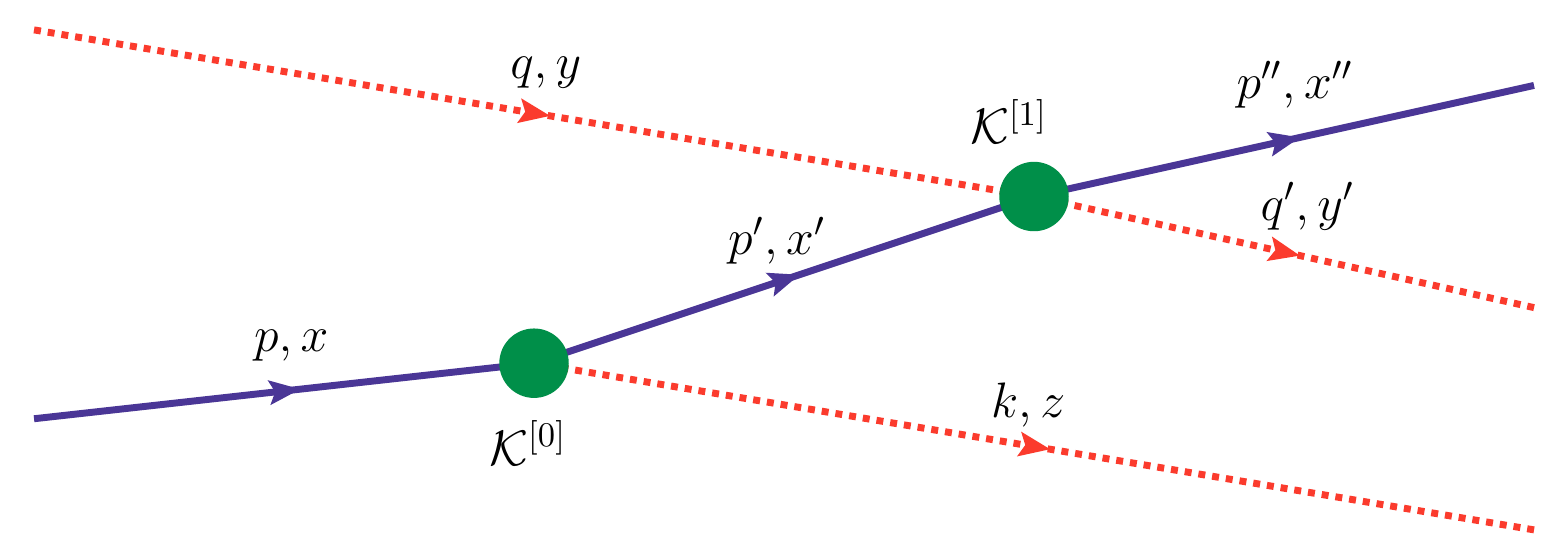}
\caption{Schematics of a pion decaying into a soft and a hard photon, with the hard photon ultimately detected through an interaction
in which it exchanges a small part of its momentum
with a particle in a detector (hard worldlines in solid blue,
soft worldlines in dotted red)}
\label{pic:kbobprl}
\end{center}
\end{figure}

In relation to the discussion we offered in the previous two sections,
an important observation reported in Ref.~\cite{anatomy}
concerns the fact that in the relative-locality framework
 a particle
is still ``essentially free" when its interactions only involve
exchanges of very small fractions of its momentum.
So in this limit applications of the relative-locality framework to
the case of a $\kappa$-Poincar\'e-inspired momentum space
should reproduce the results we summarized in the previous two sections.
This is indeed what was found in  Ref.~\cite{anatomy},
following a strategy of analysis that we shall now revisit.
One can consider the situation shown
in Figure~\ref{pic:kbobprl}.

To give some tangibility  to the situation shown in Figure~\ref{pic:kbobprl}
one could view, for example, the incoming blue worldline $p,x$
as a highly boosted pion, which decays  producing
a hard photon ($p',x'$) and a very soft photon ($k,z$);
then the hard photon propagates freely until it
it exchanges a small amount of momentum with
a soft particle ($q,y$).
One can then ask if and how, for fixed time of emission (first interaction),
 the time of detection (second interaction)
of the hard photon
depends on its momentum $p'$.

An action which is suitable for the relative-locality-framework description
of  the process shown in Figure~\ref{pic:kbobprl} is~\cite{anatomy}
 \begin{equation}
\begin{split}
\mathcal{S}^{\kappa (2)} = & \int_{s_{0}}^{+\infty}ds\left(z^{\mu}\dot{k}_{\mu}-\ell z^{1}k_{1}\dot{k}_{0}+\mathcal{N}_k\mathcal{C}_\kappa\left[k\right]\right)+
\int_{-\infty}^{s_{0}}ds\left({x}^{\mu}\dot{p}_{\mu}-\ell x^{1}p_{1}\dot{p}_{0}+\mathcal{N}_p\mathcal{C}_\kappa\left[p\right]\right)\\
  + &\int_{s_{0}}^{s_{1}}\!\!
ds\left({x'}^{\mu}\dot{p'}_{\mu}-\ell{x'}^{1}p'_{1}\dot{p}'_0+\mathcal{N}_{p'}\mathcal{C}_\kappa\left[p'\right]\right)+\int_{s_{1}}^{+\infty}\!\!ds\left(x''^{\mu}\dot{p''}_{\mu}-\ell x''^{1}p''_{1}\dot{p''}_{0}+\mathcal{N}_{p''}\mathcal{C}_\kappa\left[p''\right]\right)\\
 &+\int_{-\infty}^{s_{1}}ds\left(y^{\mu}\dot{q}_{\mu}-\ell y^{1}q_{1}\dot{q}_{0}+\mathcal{N}_q\mathcal{C}_\kappa\left[q\right]\right) +\int_{s_{1}}^{+\infty}ds\left(y'^{\mu}\dot{q'}_{\mu}-\ell {y}'^{1}{q}_{1}'{\dot{q}}_{0}'+\mathcal{N}_{q'}\mathcal{C}_\kappa\left[{q'}\right]\right)\\
 & -\xi_{[0]}^{\mu}\mathcal{K}_{\mu}^{[0]}(s_{0})
 -\xi_{[1]}^{\mu}\mathcal{K}_{\mu}^{[1]}(s_{1})\ ,
\end{split}
\label{action k-bob}
\end{equation}
where
\begin{gather}\nonumber
{\cal K}^{[0]}_\mu (s_{0})= (q \oplus p)_\mu-(q\oplus p'\oplus k)_\mu = p_\mu - p_\mu' - k_\mu - \ell \delta_\mu^1 ( - q_0 p_1 + q_0 p_1' + q_0 k_1' +p_0'k_1) \ ,\\
\begin{split}
{\cal K}^{[1]}_\mu (s_{1}) & =  (q\oplus p'\oplus k)_\mu - (p''\oplus q'\oplus k)_\mu \\& 
= q_\mu +p_\mu' -p_\mu''-q_\mu'-\ell \delta_\mu^1 (-q_0p_1' - q_0k_1 - p_0'k_1 +p_0''q_1' + p_0''k_1 + q_0'k_1) \ .
\end{split}
\label{constraints}
\end{gather}

From this action one obtains easily the
 equations of motion and the constraints,
\begin{gather*}
\dot p_\mu =0~,~~\dot q_\mu =0~,~~\dot k_\mu =0~,~~\dot p'_\mu =0~,~~\dot p''_\mu =0\ ,\\
\mathcal{C}_{\kappa}[p]=0~,~~\mathcal{C}_{\kappa}[q]=0~,~~\mathcal{C}_{\kappa}[k]=0~,~~\mathcal{C}_{\kappa}[p']=0~,~~\mathcal{C}_{\kappa}[p'']=0\ ,\\
\dot z^\mu - \mathcal{N}_k \left(\frac{\delta \mathcal{C}_{\kappa}[k]}{\delta k_\mu} +\ell \delta^\mu_0 \frac{\delta \mathcal{C}_{\kappa}[k]}{\delta k_1} k_1\right)=0~,~~~
\dot y^\mu - \mathcal{N}_q \left(\frac{\delta \mathcal{C}_{\kappa}[q]}{\delta q_\mu} +\ell \delta^\mu_0 \frac{\delta \mathcal{C}_{\kappa}[q]}{\delta q_1} q_1\right)=0~,~~~\\
{\dot y}'^\mu - \mathcal{N}_{q'} \left(\frac{\delta \mathcal{C}_{\kappa}[q']}{\delta q_\mu'} +\ell \delta^\mu_0 \frac{\delta \mathcal{C}_{\kappa}[q']}{\delta q_1'} q_1'\right)=0~,~~~
\dot x^\mu - \mathcal{N}_p \left(\frac{\delta \mathcal{C}_{\kappa}[p]}{\delta p_\mu} +\ell \delta^\mu_0 \frac{\delta \mathcal{C}_{\kappa}[p]}{\delta p_1} p_1\right)=0~,~~~\\
\dot x'^\mu - \mathcal{N}_{p'} \left(\frac{\delta \mathcal{C}_{\kappa}[p']}{\delta p'_\mu} +\ell \delta^\mu_0 \frac{\delta \mathcal{C}_{\kappa}[p']}{\delta p'_1} p'_1\right)=0~,~~~
\dot x''^\mu - \mathcal{N}_{p''} \left(\frac{\delta \mathcal{C}_{\kappa}[p'']}{\delta p''_\mu} +\ell \delta^\mu_0 \frac{\delta \mathcal{C}_{\kappa}[k]}{\delta p''_1} p''_1\right)=0~,~~~
\end{gather*}
and the boundary conditions:
\begin{gather}
z^\mu(s_{0}) = -\xi^\nu_{[0]} \left(\frac{\delta \mathcal{K}^{[0]}_\nu}{\delta k_\mu}+\ell \delta^\mu_0 \frac{\delta \mathcal{K}^{[0]}_\nu}{\delta k_1}k_1\right)\nonumber\ ,\\
x^\mu(s_{0}) = \xi^\nu_{[0]} \left(\frac{\delta \mathcal{K}^{[0]}_\nu}{\delta p_\mu}+\ell \delta^\mu_0 \frac{\delta \mathcal{K}^{[0]}_\nu}{\delta p_1}p_1\right)\nonumber\ ,\\
x'^\mu(s_{0}) = -\xi^\nu_{[0]} \left(\frac{\delta \mathcal{K}^{[0]}_\nu}{\delta p'_\mu}+\ell \delta^\mu_0 \frac{\delta \mathcal{K}^{[0]}_\nu}{\delta p'_1}p'_1\right)\ , \qquad x'^\mu(s_{1}) = \xi^\nu_{[1]} \left(\frac{\delta \mathcal{K}^{[1]}_\nu}{\delta p'_\mu}+\ell \delta^\mu_0 \frac{\delta \mathcal{K}^{[1]}_\nu}{\delta p'_1}p'_1\right)\nonumber\ ,\\
x''^\mu(s_{1}) = -\xi^\nu_{[1]} \left(\frac{\delta \mathcal{K}^{[1]}_\nu}{\delta p''_\mu}+\ell \delta^\mu_0 \frac{\delta \mathcal{K}^{[1]}_\nu}{\delta p''_1}p''_1\right)\nonumber\ ,\\
y^\mu(s_{1})= \xi^\nu_{[1]} \left(\frac{\delta \mathcal{K}^{[1]}_\nu}{\delta q_\mu}+\ell \delta^\mu_0 \frac{\delta \mathcal{K}^{[1]}_\nu}{\delta q_1}q_1\right)\nonumber\ ,\\
y'^{\mu} (s_{1}) = -\xi^\nu_{[1]} \left(\frac{\delta \mathcal{K}^{[1]}_\nu}{\delta q_\mu'}+\ell \delta^\mu_0 \frac{\delta \mathcal{K}^{[1]}_\nu}{\delta q_1'}q_1'\right)\nonumber \ .
\end{gather}

Evidently here the issue of interest is primarily contained in the
dependence of the time of detection at a given detector
of simultaneously-emitted particles on
the momenta of the particles and on the specific properties
of the interactions involved in the analysis.
We start by noticing that
for the particle of worldline $x^\mu$, we have
\begin{equation}
x^1 (s) =x^1 (\bar{s})+ v^1(x^0(s)-x^0(\bar{s}))  \label{worldlinek-bob}\ ,
\end{equation}
which in the massless case
(and whenever  $m/{p_1}^2 \ll |\ell p_1|$) takes the simple form
\begin{equation}
 x^1 (s) =x^1 (\bar s)- \frac{p_1}{|p_1|}(x^0 (s) - x^0 (\bar s))\ .
\label{speedmassless}
\end{equation}
In obtaining (\ref{speedmassless}) we used the on-shell relation
\begin{gather*}
p_0=\sqrt{p_1^2 + m^2}-\frac{\ell}{2} p_1^2 \ ,
\end{gather*}
and the fact that for $m/{p_1}^2 \ll |\ell p_1|$
(consistently again with our choice of conventions, which is such
that $v^1>0 \Longrightarrow p^1 <0$)
\begin{gather*}
v^1=\frac{\dot x^1}{\dot x^0}=-\frac{p_1}{p_0}(1-\ell p_0 +\frac{\ell p_1^2}{2 p_0})=-\frac{p_1}{|p_1|} ~.
\end{gather*}

As in the previous section we have here momentum-independent coordinate speeds for
massless particles, so in particular according to Alice's coordinates
two massless particles of momenta $p_1^s$ and $p_1^h$
simultaneously emitted at Alice (in Alice's spacetime origin)
appear to reach detector Bob simultaneously, apparently establishing
a coincidence of detection events.
But once again the presence of relative locality
evidently requires that
in order to establish
 the dependence of the time of detection
on the momentum of the massless particles we must again transform
the relevant
worldlines to the corresponding description
by an observer Bob local to the detection.
Let us then return to the two-interaction process of Fig.~\ref{pic:kbobprl}
and take as our hard massless particle of momentum $p_1^h$
the particle in that process which we had originally labeled as having
momentum $p'_1$.
For the process of Fig.~\ref{pic:kbobprl}
the description of the transformation from Alice's to Bob's worldlines is~\cite{anatomy}
\begin{equation}
\begin{split}
z_{B}^{0}(s)&=  z_{A}^{0}(s)+b^{\mu}  \{ (q\oplus p'\oplus k)_\mu , z^0\} = z_{A}^{0}(s)-b^{0}-\ell b^{1}k_{1} \simeq  z_{A}^{0}(s)-b^{0}\ ,\\
z_{B}^{1}(s)&=z_{A}^{1}(s)+b^{\mu}  \{ (q\oplus p'\oplus k)_\mu , z^1\}
={z}_{A}^{1}(s)-b^{1} -\ell(p_0'+q_0) \simeq {z}_{A}^{1}(s)-b^{1} -\ell b^1 p_0' \ ,\\
x_{B}^{0}(s)&=  x_{A}^{0}(s)+b^{\mu}  \{ (q \oplus p)_\mu , x^0\}  =x_{A}^{0}(s)-b^{0}-\ell b^{1}p_{1} \ ,\\
x_{B}^{1}(s)&=x_{A}^{1}(s)+b^{\mu}  \{ (q \oplus p)_\mu , x^1\}  ={x}_{A}^{1}(s)-b^{1}-\ell q_0  \simeq {x}_{A}^{1}(s)-b^{1} \ ,\\
{x'}_{B}^{0}(s)&=  {x'}_{A}^{0}(s)+b^{\mu}  \{ (q\oplus p'\oplus k)_\mu , x'^0\}
={x'}_{A}^{0}(s)-b^{0}-\ell b^{1}(k_{1}+p'_1) \simeq {x'}_{A}^{0}(s)-b^{0}-\ell b^{1}p'_{1}\ ,\\
{x'}_{B}^{1}(s)&=x_{A}^{1}(s)+b^{\mu}  \{ (q\oplus p'\oplus k)_\mu , x'^1\}
={x'}_{A}^{1}(s)-b^{1}-\ell q_0 \simeq {x'}_{A}^{1}(s)-b^{1}\ , \\
{x''}_{B}^{0}(s)&=\!{x''}_{A}^{0}(s)\!+\!b^{\mu} \{ (p''\!\oplus\! q'\!\oplus\! k\!)_\mu , x''^0\}\!
=\!{x''}_{A}^{0}(s)\!-\!b^{0}\!-\!\ell b^{1}\!(q_1'\!+\!k_1\!+\! p''_{1})\! \simeq\! {x''}_{A}^{0}(s)\!-\!b^{0}\!-\!\ell b^{1}p''_{1}\ \!\!,\\
{x''}_{B}^{1}(s)&={x''}_{A}^{1}(s)+b^{\mu} \{ (p''\oplus q'\oplus k)_\mu , x''^1\}  ={x''}_{A}^{1}(s)-b^{1}\ ,\\
y_{B}^{0}(s)&=  y_{A}^{0}(s)+b^{\mu}  \{ (q\oplus p'\oplus k)_\mu , y^0\}  =y_{A}^{0}(s)-b^{0}-\ell b^{1}(p'_1+k_1+q_1) \simeq y_{A}^{0}(s)-b^{0}-\ell b^{1}p'_1\ ,\\
y_{B}^{1}(s)&=y_{A}^{1}(s)+b^{\mu}  \{ (q\oplus p'\oplus k)_\mu , y^1\}  ={y}_{A}^{1}(s)-b^{1}\ ,\\
{y'}_{B}^{0}(s)&=  {y'}_{A}^{0}(s)+b^{\mu} \{ (p''\oplus q'\oplus k)_\mu , y'^0\} ={y'_{A}}^{0}(s)-b^{0}-\ell b^{1}(k_1+q'_1) \simeq {y'_{A}}^{0}(s)-b^{0} \ ,\\
{y'}_{B}^{1}(s)&={y'}_{A}^{1}(s)+b^{\mu} \{ (p''\oplus q'\oplus k)_\mu , y'^1\} ={y'_{A}}^{1}(s)-b^{1} - \ell b^1 p_0''\ .
\end{split}
\label{translationsk-bob}
\end{equation}
Using these transformation laws it is easy to recognize that, having dropped
the negligible ``soft terms" from small momenta,
indeed we are obtaining results that are fully consistent
with the ones obtained in the Hamiltonian description of free particles.
To see this explicitly let us
consider the situation where, simultaneously to the interaction emitting
the hard particle $x',p'$ in Alice origin, we also have the emission of
a soft photon $x_s, p_s$.\\
And as observer Bob let us take one who
is reached {\underline{in its spacetime origin}}
 by the soft photon emitted by Alice. For the event of detection
 of the hard particle  $x',p'$
 we take one such that it occurs in Bob's {\underline{spatial origin}}.\\
 From a relative-locality perspective the setup we are arranging is
 such that ``Alice is an emitter" (the spatial origin of Alice's coordinate system
 is an ideally compact,
 infinitely small, emitter) and ``Bob is a detector"
(the spatial origin of Bob's coordinate system
 is an ideally compact,
 infinitely small, detector). The two worldlines we focus on, a soft and
 a hard worldline,
 both originate from Alice's spacetime origin (they are both emitted by Alice,
 in the spatial origin of Alice's frame of reference, and both
 at time $t_{Alice}=0$)
 and both end up being detected by Bob, but, while by construction the soft
 particle reaches Bob's spacetime origin, the time at which the hard particle
 reaches Bob spatial origin is to be determined.

Reasoning  as usual
at first order in $\ell$, it is easy to verify that Bob describes
the ``interaction coordinate" ${\xi_B^{[1]}}^\mu$ of the interaction at $s=s_1$
as coincident with the $s=s_1$ endpoints of the
worldlines $x',p'$; $x'',p''$; $q,y$; $q',y'$:
\begin{equation}
{\xi_B^{[1]}}^\mu = {x_B'}^\mu(s_{1}) = {x_B''}^\mu(s_{1}) = y_B^\mu(s_{1}) = {y_B'}^\mu(s_{1}) \ .
 \label{delaycoordPRE}
\end{equation}
We take into account that there are no relative-locality effects
in the description given by Bob whenever an interaction occurs ``in the vicinity of Bob":
our leading-order analysis assumes the available sensitivity
is sufficient to expose manifestations of relativity of locality
of order $\ell p_h L$ (where $L$ is the distance from the interaction-event
to the origin
of the observer and $p_h$ is a ``suitably high" momentum),
with $L$ set in this case by the distance Alice-Bob, so even a hard-particle
interaction which is at a distance $d$ from the origin of Bob
will be treated as absolutely local by Bob if $d \ll L$.\\
According to this both ``detection events" are absolutely local
for observer Bob: of course this is true for
the  event of detection of the soft photon $x_s, p_s$
(which we did not even specify since its softness ensures us of its
absolute locality) and it is also true for the interaction-event
of ``detection near Bob"
of the hard particle $x',p'$. Ultimately this allows us
 to handle the time component of the coordinate fourvector (\ref{delaycoordPRE})
  as the actual delay that Bob measures
between the two detection times:
\begin{equation}
 \Delta t = {\xi_B^{[1]}}^0 = {x_B'}^0(s_{1}) = {x_B''}^0(s_{1}) = y_B^0(s_{1}) = {y_B'}^0(s_{1}) \label{delaycoord}\ .
\end{equation}

\begin{figure}[h!]
\begin{center}
\includegraphics[width=0.44 \textwidth]{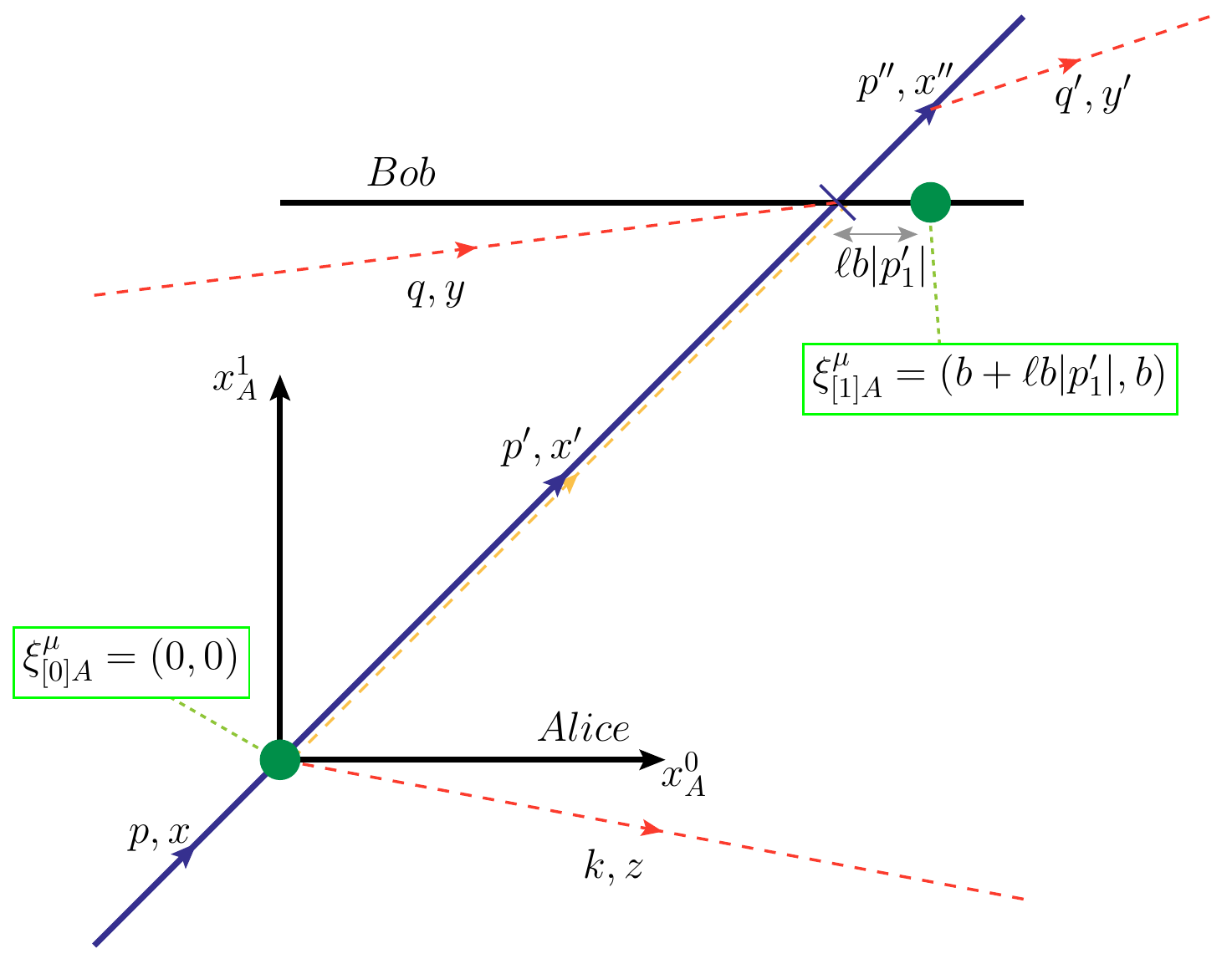}
\includegraphics[width=0.44 \textwidth]{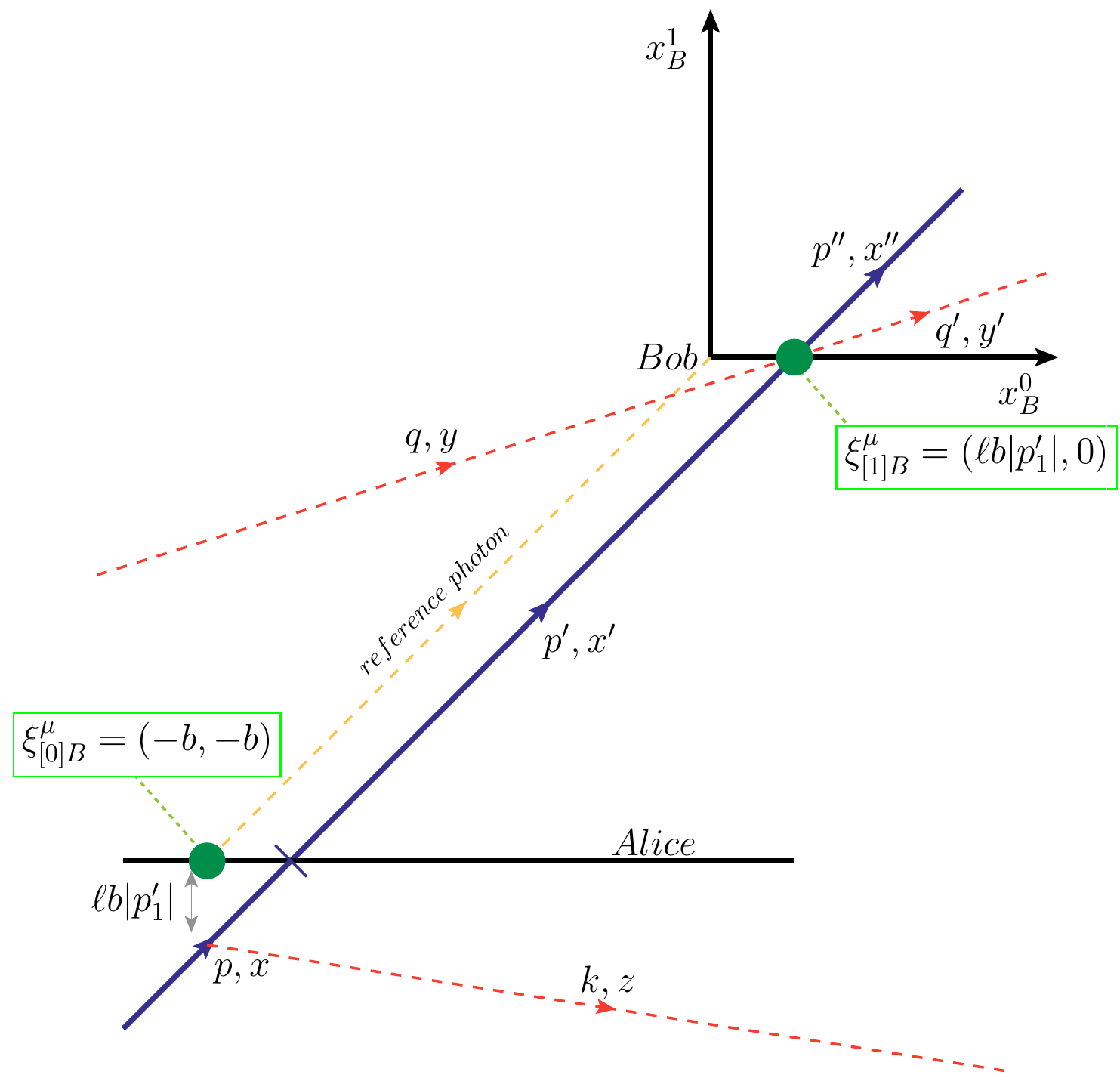}
\caption{Schematic description of the time delay discussed in this subsection.
The various agents in the analysis are shown both as described by Alice
(left panel) and as described by Bob (right panel). These are spacetime graphs
(in a 2D spacetime) showing the actual worldlines of particles. In addition to the
two hard interactions we are considering
(qualitatively described already in Figure~\ref{pic:kbobprl}),
we also show (as the orange-dotted worldline) a soft photon going from Alice's
origin to Bob's origin. As shown in the figure the derivation is arranged 
 so that all emissions and detections occur in the spatial origin
of either Alice or Bob (but because of the relative locality according to Alice
the hard detections at/near Bob would be nonlocal interactions and according to Bob
the hard emissions at Alice would be nonlocal processes). We also show,
as the bulky green dots, the results obtained in Ref.~\cite{anatomy} for
the formal positions of the interaction points, as coded in the
formal ``interaction coordinates" $\xi^\mu$ .}
\label{alicebobkappabobprl}
\end{center}
\end{figure}

From the equations (\ref{translationsk-bob}) relative to the worldline $x',p'$, it follows that
\begin{equation}
 {x'_A}^1(s_{1}) = {x'_B}^1(s_{1}) + b^1 = b^1\ ,
\end{equation}
from which, considering the worldlines (\ref{speedmassless}),
 it follows that
  (assuming indeed $m/(p'_1)^2 \ll |\ell p'_1|$)
 Alice ``sees'' the $s=s_{1}$ endpoint of the worldline $x',p'$ at the coordinates
\begin{equation}
{x'}^\mu_A(s_{1}) = {x'}^\mu_B(s_{1}) + b^\mu = b^\mu = (b,b)\ .
\end{equation}
And then, from the equations (\ref{translationsk-bob}) and (\ref{delaycoord}), it follows that Bob measures the delay~\cite{anatomy}
\begin{equation}
\Delta t = {x_B'}^0(s_{1}) = {x'}_{A}^{0}(s_{1})-b^{0}-\ell b^{1}p'_{1} = \ell b |p'_{1}| \ ,
\label{delay}
\end{equation}
in agreement with the results we summarized in the previous two sections.
 The strategy of this analysis is schematically reported in Figure \ref{alicebobkappabobprl}.

\subsection{Role of ``hard interactions"}\label{secsuperpion}
The results of Ref.~\cite{anatomy} which we summarized in the previous subsection
concerned the limit in which a particle exchanged among interactions 
propagates ``essentially free", originating from an interaction in which it
is the only outgoing hard particle and ending into an interaction in which
it is the only incoming hard particle.
We shall now summarize the corresponding results of Ref.~\cite{anatomy} which concern
 more general cases.
Our main objective in this subsection is to highlight the results of Ref.~\cite{anatomy}
which establish a possible dependence of certain aspects of the propagation of a particle
on the actual emission and detection interaction which that particle connects.

For these purposes it suffices to modify the analysis summarized in
the previous subsection
in rather minor way: the new features become evident when~\cite{anatomy} 
at least one of the interactions (emission and/or detection)
involves at least 3 hard particles in total, among in and out particles.
As an example of this situation, Ref.~\cite{anatomy}
considers the case
of a ultraenergetic particle at rest decaying into two particles, both hard,
one of which is the particle detected at our observatory.

\begin{figure}[H]
\begin{center}
\includegraphics[width=0.66 \textwidth]{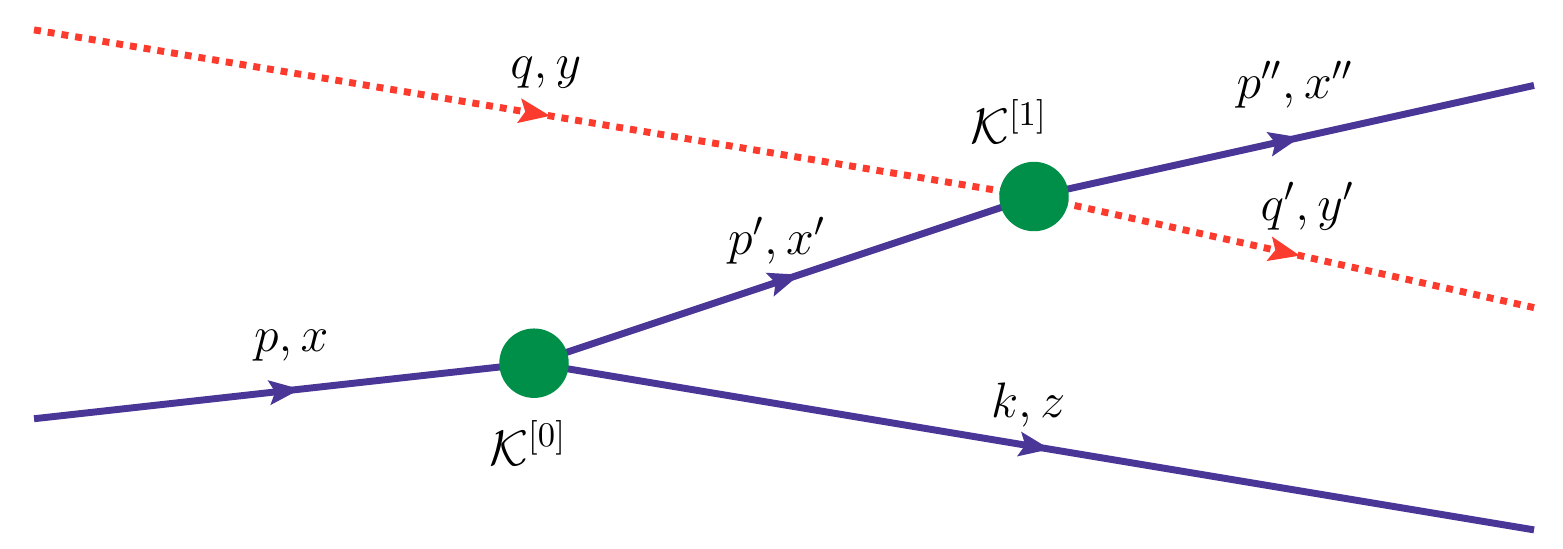}
\caption{Schematic description of a case where a hard ultrarelativistic particle
originating from a hard emission interaction
(one hard particle in, two hard particles out) is detected
in a soft interaction (only one hard particle in and only one hard particle
out). Solid-blue lines are for hard particles, dashed-red lines are for soft particles.}
\label{figsuperpion}
\end{center}
\end{figure}

As shown in Fig.~\ref{figsuperpion} the analysis
can be arranged in exactly the same way as
in the previous subsection, with a trivalent vertex for the emission
interaction and a four-valent vertex for the detection. And the kinematics at the
four-valent vertex is left unchanged, involving a soft particle in the in state
and a soft particle in the out state. What changes in this case is the kinematics of
the emission vertex, now assuming that all particles involved are hard.

In this case it is actually important to consider at least two possibilities~\cite{anatomy},
since this allows to put in focus some of the relevant issues.
The first possibility is encoded in the following choice of boundary conditions
at endpoints of worldlines:
\begin{gather}\nonumber
{\cal K}^{[0]}_\mu (s_{0})= (q \oplus p)_\mu-(q\oplus p'\oplus k)_\mu = p_\mu - p_\mu' - k_\mu - \ell \delta_\mu^1 ( - q_0 p_1 + q_0 p_1' + q_0 k_1' +p_0'k_1)\ , \\
\begin{split}
{\cal K}^{[1]}_\mu (s_{1}) & = (q\oplus p'\oplus k)_\mu - (p''\oplus q'\oplus k)_\mu \\& = q_\mu +p_\mu' -p_\mu''-q_\mu'-\ell \delta_\mu^1 (-q_0p_1' - q_0k_1 - p_0'k_1 +p_0''q_1' + p_0''k_1 + q_0'k_1) \ .
\end{split}
\end{gather}
The worldlines seen by observer/detector Bob, distant from the emission,
that follow from this choice of boundary terms
have been already given in Eq.~(\ref{speedmassless}).
The main difference between the situation in the previous
subsection and this situation is that
the ``primary", the  particle incoming to the emission interaction,
is at rest, with $p_{1}=0$, which also implies that the two outgoing particles
of the emission interaction must both be hard.
 For the worldlines involved in the emission interaction this leads to
\begin{equation}
\begin{split}
x_{B}^{0}(s)&=x_{A}^{0}(s)+b^{\mu}  \{ (q \oplus p)_\mu , x^0\} =x_{A}^{0}(s)-b^{0}-\ell b^{1}p_{1} =x_{A}^{0}(s)-b^{0}\ ,\\
x_{B}^{1}(s)&=x_{A}^{1}(s)+b^{\mu}  \{ (q \oplus p)_\mu , x^1\}={x}_{A}^{1}(s)-b^{1}-\ell q_0\simeq {x}_{A}^{1}(s)-b^{1}\ ,\\
{x'}_{B}^{0}(s)&={x'}_{A}^{0}(s)+b^{\mu}  \{ (q\oplus p'\oplus k)_\mu , x'^0\}={x'}_{A}^{0}(s)-b^{0}-\ell b^{1}(k_{1}+p'_1)={x'}_{A}^{0}(s)-b^{0}\ ,\\
{x'}_{B}^{1}(s)&=x_{A}^{1}(s)+b^{\mu}  \{ (q\oplus p'\oplus k)_\mu , x'^1\} ={x'}_{A}^{1}(s)-b^{1}-\ell q_0\simeq{x'}_{A}^{1}(s)-b^{1}\ ,\\
z_{B}^{0}(s)&=z_{A}^{0}(s)+b^{\mu}  \{ (q\oplus p'\oplus k)_\mu , z^0\}= z_{A}^{0}(s)-b^{0}-\ell b^{1}k_{1} \ ,\\
z_{B}^{1}(s)&=z_{A}^{1}(s)+b^{\mu}  \{ (q\oplus p'\oplus k)_\mu , z^1\}={z}_{A}^{1}(s)-b^{1} -\ell(p_0'+q_0) \simeq {z}_{A}^{1}(s)-b^{1} -\ell b^1 p_0' \ .
\end{split}
\label{translationsnodelay}
\end{equation}

And from this one easily sees that the particle $p',x'$, the particle
then detected at Bob, translates classically, without any deformation term.\\
{\underline{So this time we have that no momentum dependence of the times
of arrival is predicted}}
$$t_{detection}= x_{B}'^{0}(s_{1})=x_{A}'^{0}(s_{1})-b^{0}=0~.$$

As announced, in this case one must however consider at least one alternative possibility
for the choice of boundary conditions.
As observed in Ref.~\cite{anatomy} in cases, like the emission interaction
we are presently contemplating, in which an interaction involves only hard
particles the noncommutativity of the composition law
can play a highly non-trivial role. As a result there is particular interest
in analyzing~\cite{anatomy}
the following alternative choice of ${\cal K}$'s for the boundary terms
\begin{gather}\nonumber
{\cal K}^{[0]}_\mu = (p \oplus q)_\mu-(k\oplus  p' \oplus q)_\mu = p_\mu - p_\mu' - k_\mu - \ell \delta_\mu^1 ( - p_0 q_1 + k_0 p_1' + k_0 q_1 +p_0' q_1)\ , \\
\begin{split}
{\cal K}^{[1]}_\mu & = (k\oplus  p' \oplus q)_\mu - (k \oplus  p'' \oplus q')_\mu \\& = p_\mu' +q_\mu -p_\mu''-q_\mu'-\ell \delta_\mu^1 (-k_0p_1' - k_0 q_1 - p_0' q_1 +k_0 p_1'' + k_0 q_1' + p_0'' q_1' ) \ .
\end{split}
\end{gather}
Focusing again on the worldline $x' , p'$ detected at Bob
one then finds~\cite{anatomy}
\begin{equation}
\begin{split}
{x'}_{B}^{0}(s)&={x'}_{A}^{0}(s)+b^{\mu}  \{ (k\oplus  p' \oplus q)_\mu , x'^0\}={x'}_{A}^{0}(s)-b^{0}-\ell b^{1}(q_{1}+p'_1) \simeq {x'}_{A}^{0}(s)-b^{0}-\ell b^{1}p'_{1}\ ,\\
{x'}_{B}^{1}(s)&=x_{A}^{1}(s)+b^{\mu}  \{ (k\oplus  p' \oplus q)_\mu , x'^1\} ={x'}_{A}^{1}(s)-b^{1}-\ell b^{1}k_{0} \ .
\end{split}
\end{equation}
And from the equation of motion (\ref{speedmassless})
one now deduces that
$${x'}_{B}^{1}(s)={x'}_{B}^{0}(s)-\ell b^{1}(k_{0}-p'_{1}) \ ,$$
which in turn implies that the time of detection at Bob
of the particle with worldline $x' , p'$
is~\cite{anatomy}
\begin{equation}
 t_{detection} = {x_B'}^0(s_{1})  =-\ell b^{1}(p'_{1}-k_{0})=2\ell b^{1}|p'_{1}| \ .
\end{equation}
{\underline{The \!dependence of the \!time \!of \!detection on \!the \!momentum of
\!the \!particle being \!detected is \!back!}}\\
{\underline{And this dependence is twice as \!strong \!as the dependence
on momentum found in the previous}}\\ {\underline {subsection!}}

So these results of Ref.~\cite{anatomy} expose, within the relative-locality framework, 
a rather striking aspect of the dependence of the propagation of a particle
on the actual emission and detection interaction which that particle connects.
For what concerns
times of detection of simultaneously emitted massless particles of  momentum $p'_1$,
emitted from a source at a distance $L$ from the detector one encounters
3 situations:\\
{\bf (case A)} the emission interaction involves only one hard incoming particle
and one hard outgoing particle, all other particles in the emission
interaction being soft: \\
the times of arrival have a dependence
on momentum governed by
$$t_{detection} = \ell L |p'_{1}|$$
and this result is independent of the position occupied by the
momentum $p'_\mu$ in our noncommutative composition law\\
{\bf (case B)} the emission interaction is the decay of a ultra-high-energy
 particle at rest, involves a total of 3 hard particles,
and the momentum $p'_\mu$ appears in the composition of momenta
to the {\underline{left}} of a hard particle: \\
the times of arrival have no dependence
on momentum
$$t_{detection} = 0$$
{\bf (case C)} the emission interaction is the decay of a ultra-high-energy
 particle at rest, involves a total of 3 hard particles,
and the momentum $p'_\mu$ appears in the composition of momenta
to the {\underline{right}} of a hard particle: \\
the times of arrival have the following dependence
on momentum
$$t_{detection} = 2 \ell L |p'_{1}|$$
(twice as large as in the case A).

\section{Closing remarks}
We have here summarized some of the recent results~\cite{bob,k-bob,anatomy} which further
contributed to establishing the possibility of relying
on $\kappa$-Poincar\'e phase spaces for the description
of a momentum dependence of the speed of massless particles.
Awareness of the possibility of a relativity of spacetime locality
evidently proves
to be a key resource from this perspective. And the recently-proposed
relative-locality framework has remarkable potentialities 
for further empowering this approach (and many other approaches).
It will be interesting to explore the implications of a possible
relativity of spacetime locality for other scenarios
being considered from a DSR/deformed-Lorentz-symmetry perspective
(see, {\it e.g.}, Refs.~\cite{jurekdsrREVIEW,gacdsrrev2010,dsrgzk,dsrPOLAND2001}).

The fact that effects of the type here discussed are subjectable
to experimental testing, even if $|\ell|^{-1}$ is as big as the Planck 
scale~\cite{gacLRR,grbgac,gampul,schaefer,fermiSCIENCE,ellisUNO,gacsmolinPRD,fermiNATURE}, 
must remain the main goal of this research programme, even though it
is amusing that in ongoing analyses of the anomaly tentatively reported
by the OPERA collaboration~\cite{opera}
some of the concepts that were here relevant 
are contributing to valuable clarifications~\cite{whataboutopera,operaDSR,synchroDSR}.

Specifically for what concerns $\kappa$-Poincar\'e-inspired research
future studies should give high priority to the exploration of formulations
that are alternative to the one adopted in Refs.~\cite{bob,k-bob,anatomy},
which are here summarized.
Refs.~\cite{bob,k-bob,anatomy} all
worked throughout consistently inspired by the so-called time-to-the-right ordering
convention~\cite{majrue}, 
which is by far the choice most frequently preferred in the literature.
But a lot remains to be understood concerning the relationship
between this ordering prescription
and other ordering prescription, such as the ones 
discussed in Refs.~\cite{gacAlessandraFrancesco,meljaEPJ}.

\end{document}